\documentclass[12pt,english]{article}
\usepackage{times}
\usepackage[T1]{fontenc}
\usepackage{geometry}
\usepackage{graphicx}
\usepackage{setspace}
\usepackage{amssymb}

\makeatletter

\providecommand{\tabularnewline}{\\}

\usepackage{bm}

\usepackage{babel}
\makeatother
\begin{document}

\section*{On the Design of Complex \emph{EM} Devices and Systems through the
System-by-Design Paradigm - A Framework for Dealing with the Computational
Complexity}

\noindent \vfill

\noindent A. Massa,$^{(1)(2)(3)(4)}$ \emph{Fellow, IEEE}, and \emph{}M.
Salucci,$^{(2)(4)}$ \emph{Member, IEEE}

\noindent \vfill

\noindent {\small $^{(1)}$} \emph{\small ELEDIA Research Center}
{\small (}\emph{\small ELEDIA}{\small @}\emph{\small UESTC} {\small -
UESTC)}{\small \par}

\noindent {\small School of Electronic Engineering, Chengdu 611731
- China}{\small \par}

\noindent \textit{\emph{\small E-mail:}} \emph{\small andrea.massa@uestc.edu.cn}{\small \par}

\noindent {\small Website:} \emph{\small www.eledia.org/eledia}{\small -}\emph{\small uestc}{\small \par}

\noindent {\small ~}{\small \par}

\noindent {\small $^{(2)}$} \emph{\small ELEDIA Research Center}
{\small (}\emph{\small ELEDIA}{\small @}\emph{\small UniTN} {\small -
University of Trento)}{\small \par}

\noindent {\small Via Sommarive 9, 38123 Trento - Italy}{\small \par}

\noindent \textit{\emph{\small E-mail:}} {\small \{}\emph{\small andrea.massa}{\small ,}
\emph{\small marco.salucci}{\small \}@}\emph{\small unitn.it}{\small \par}

\noindent {\small Website:} \emph{\small www.eledia.org/eledia-unitn}{\small \par}

\noindent {\small ~}{\small \par}

\noindent {\small $^{(3)}$} \emph{\small ELEDIA Research Center}
{\small (}\emph{\small ELEDIA@TSINGHUA} {\small - Tsinghua University)}{\small \par}

\noindent {\small 30 Shuangqing Rd, 100084 Haidian, Beijing - China}{\small \par}

\noindent {\small E-mail:} \emph{\small andrea.massa@tsinghua.edu.cn}{\small \par}

\noindent {\small Website:} \emph{\small www.eledia.org/eledia-tsinghua}{\small \par}

\noindent {\small ~}{\small \par}

\noindent {\small $^{(4)}$} \emph{\small ELEDIA Research Center}
{\small (}\emph{\small ELEDIA}{\small @}\emph{\small L2S} {\small -
UMR 8506)}{\small \par}

\noindent {\small 3 rue Joliot Curie, 91192 Gif-sur-Yvette - France}{\small \par}

\noindent \textit{\emph{\small E-mail:}} {\small \{}\emph{\small andrea.massa}{\small ,}
\emph{\small marco.salucci}{\small \}}\emph{\small @l2s.centralesupelec.fr}{\small \par}

\noindent {\small Website:} \emph{\small www.eledia.org/eledia-l2s}{\small \par}

\vfill

\emph{This work has been submitted to the IEEE for possible publication.
Copyright may be transferred without notice, after which this version
may no longer be accessible.}

\noindent \vfill

\newpage
\section*{On the Design of Complex \emph{EM} Devices and Systems through the
System-by-Design Paradigm - A Framework for Dealing with the Computational
Complexity}

\vfill

\begin{flushleft}A. Massa and \emph{}M. Salucci\end{flushleft}

\noindent \vfill

\begin{abstract}
\noindent {\footnotesize The System-by-Design (}\emph{\footnotesize SbD}{\footnotesize )
is an emerging engineering framework for the optimization-driven design
of} \emph{\footnotesize complex} {\footnotesize electromagnetic (}\emph{\footnotesize EM}{\footnotesize )
devices and systems. More specifically, the computational complexity
of the design problem at hand is addressed by means of a suitable
selection and integration of} \emph{\footnotesize functional blocks}
{\footnotesize comprising problem-dependent and computationally-efficient
modeling and analysis tools as well as reliable prediction and optimization
strategies. Thanks to the suitable re-formulation of the problem at
hand as an optimization one, the profitable minimum-size coding of
the degrees-of-freedom (}\emph{\footnotesize DoFs}{\footnotesize ),
the {}``smart'' replacement of expensive full-wave (}\emph{\footnotesize FW}{\footnotesize )
simulators with proper} \emph{\footnotesize surrogate models} {\footnotesize (}\emph{\footnotesize SM}{\footnotesize s),
which yield fast yet accurate predictions starting from minimum size/reduced}
\emph{\footnotesize CPU}{\footnotesize -costs training sets, a favorable
{}``environment'' for an optimal exploitation of the features of
global optimization tools in sampling wide/complex/nonlinear solution
spaces is built. This research summary is then aimed at (}\emph{\footnotesize i}{\footnotesize )
providing a comprehensive description of the} \emph{\footnotesize SbD}
{\footnotesize framework and of its pillar concepts and strategies,
(}\emph{\footnotesize ii}{\footnotesize ) giving useful guidelines
for its successful customization and application to different} \emph{\footnotesize EM}
{\footnotesize design problems characterized by different levels of
computational complexity, (}\emph{\footnotesize iii}{\footnotesize )
envisaging future trends and advances in this fascinating and high-interest
(because of its relevant and topical industrial and commercial implications)
topic. Representative benchmarks concerned with the synthesis of single
antenna devices as well as complex array systems are presented to
highlight advantages and potentialities as well as current limitations
of the} \emph{\footnotesize SbD} {\footnotesize paradigm.}{\footnotesize \par}
\end{abstract}
\noindent \vfill

\noindent \textbf{Key words}: Complex \emph{EM} Problems, Optimization,
Surrogate Modeling, Learning-by-Examples (\emph{LBE}), System-by-Design
(\emph{SbD}).

\newpage
\section{Introduction}

In the last years, there have been many and significant progresses
in the development of numerical techniques - denoted as full-wave
(\emph{FW}) solvers - for the accurate analysis of complex electromagnetic
(\emph{EM}) devices and systems (see for examples \cite{Jin 2014}-\cite{Davidson 2011}
and the reference therein). Although highly-reliable, \emph{FW} solvers
are generally time-consuming \cite{Davidson 2011} and their exploitation
to solve \emph{complex} \emph{EM} synthesis problems%
\footnote{\noindent A problem is regarded as inherently \emph{complex} if its
solution requires significant resources, whatever the algorithm used.
In other words, {}``complexity'' is a {}``measure'' of (\emph{i})
the problem dimension, (\emph{ii}) the adopted mathematical model,
as well as (\emph{iii}) the computational burden.%
} often implies {}``local'' refinements of an initial/reference solution
based on parametric sweeps and/or trial-and-error steps. Otherwise,
design strategies involving analytic/semi-analytic methods (\emph{AM}s)
(e.g., \cite{Kim 2014}\cite{Jackson 1991}) are generally less computationally
demanding, thus allowing the use of more effective and complex synthesis
strategies (e.g., global optimization or gradient-based deterministic
and iterative methods), but they may be unreliable when dealing with
high-complexity systems since they typically approximate or even neglect
nonlinear \emph{EM} phenomena that require the \emph{FW} solution
of Maxwell's equations. Moreover, \emph{AM}s cannot deal with whatever
\emph{EM} device or system, while they are generally suitable for
canonical or rather {}``simple'' structures \cite{Jackson 1991}.
Therefore the {}``holy-grail'' in synthesizing complex EM systems
is, on the one hand, to take advantage of the modeling accuracy of
\emph{FW} solvers, on the other, to exploit global optimization strategies
for finding the global optimum (or the closest one) of the cost function
that quantifies the mismatch between user-requirements and design
outcomes. As a matter of fact, global optimization strategies, based
for example on nature-inspired evolutionary algorithms (\emph{EA}s)
\cite{Goudos 2019}-\cite{Bayraktar 2013}, have been widely applied
in many \emph{EM} engineering problems since they allow an efficient
exploration of the whole solution space and, unlike deterministic
algorithms, they require neither the analytic knowledge nor the differentiation
of the cost function. Moreover, \emph{a-priori} information (e.g.,
physical requirements or already available sub-optimal solutions)
can be introduced in a straightforward manner as additional constraints
on the iterative process of selecting trial solutions \cite{Rocca 2009}.
Of course, global optimizers require the evaluation of many solutions
(typically hundreds or thousands) to ensure an effective sampling
of the solution space and to find a solution fitting all user requirements,
thus the {}``bare'' integration of a \emph{FW} solver in an iterative
optimization tool will imply unrealistic/unaffordable computational
costs. In order to overcome those issues, different approaches have
been proposed ranging from (\emph{i}) the improvement of the convergence
rate of \emph{EA}s by (\emph{i.1}) using a set of {}``good'' (i.e.,
close to the global optimum) trial solutions at the initialization
of the optimization process \cite{Bayraktar 2011}\cite{Oliveri 2011}
and/or by (\emph{i.2}) identifying a minimum set of representative
solution parameters \cite{Salucci 2018}\cite{Lizzi 2008} up to (\emph{iii})
the reduction of the time for evaluating (i.e., the computation of
the mismatch cost function) a single solution \cite{Aliakbari 2017}-\cite{Prado 2019}
also integrating suitable coarse-to-fine \emph{space mapping} strategies
\cite{Bandler 2004}\cite{Koziel 2009}.

\noindent Within this context, the System-by-Design (\emph{SbD}) recently
emerged as an innovative paradigm able to exploit such strategies
in a more integrated and seamless fashion \cite{Salucci 2019}-\cite{Oliveri 2019}.
As a matter of fact, the \emph{SbD} enables an effective, \emph{}reliable\emph{,}
and computationally-efficient use of global optimizers for addressing
complex \emph{EM} design problems, since it is aimed at the \char`\"{}\emph{task-oriented
design, definition, and integration of system components to yield
EM devices with user-desired performance having the minim costs, the
maximum scalability, and suitable reconfigurability properties}\char`\"{}.
Applications of the \emph{SbD} to the synthesis of innovative meta-materials
\cite{Salucci 2019}-\cite{Nagar 2017}, fractal antennas \cite{Salucci 2019b},
electrically-large airborne radomes \cite{Massa 2018}, wide angle
impedance matching layers (\emph{WAIM}s) \cite{Oliveri 2015}\cite{Oliveri 2017},
and reflectarray antennas \cite{Oliveri 2019} have been recently
documented. 

\noindent The aim of this work is (\emph{i}) to provide a comprehensive
description of the \emph{SbD} framework and of its pillar concepts
and strategies, (\emph{ii}) to give useful guidelines for its successful
customization and application to different \emph{EM} design problems
that share the common issue of the computational complexity, (\emph{iii})
to envisage future trends and advances in this fascinating and high-interest
(because its relevant and topical industrial and commercial implications)
topic. 

\noindent The outline of the paper is as follows. The general principles
of the \emph{SbD} are pointed out in Sect. II, while the description
of the \emph{functional blocks} of the \emph{SbD} is given in Sect.
III. Two novel advanced \emph{SbD}-based synthesis strategies are
presented in Sect. IV. Representative synthesis benchmarks are illustrated
(Sect. V) to show the \emph{SbD} working as well as to give some proofs
of the method effectiveness and efficiency when dealing with computational
complexity issues. Some concluding remarks are finally drawn also
envisaging future trends (Sect. VI).

\section{The \emph{SbD} Paradigm}

By denoting with $\underline{\Omega}$ ($\underline{\Omega}=\left\{ \Omega_{k};\, k=1,\,...,\, K\right\} $)
the set of $K$ descriptive parameters {[}i.e., the degrees-of-freedom
(\emph{DoF}s){]} of the design problem at hand, the goal of the \emph{SbD}
is to yield, in a \emph{reasonable} time frame%
\footnote{\noindent Clearly, the expression is rather vague and intuitive. Of
course, there is the need of quantify the meaning of {}``\emph{reasonable
time frame}'' in a rigorous mathematical way for moving from a purely
qualitative statement to a more engineer-oriented/quantitative one
(see Sects. 3-4). %
} , the setup of the \emph{DoF}s, $\underline{\Omega}^{\left(end\right)}$,
so that the corresponding cost function value, $\Phi\left\{ \underline{\Omega}^{\left(end\right)}\right\} $,
differs from that of the global optimum, $\Phi\left\{ \underline{\Omega}^{\left(opt\right)}\right\} $,
at most for a maximum deviation $\xi$\begin{equation}
\underline{\Omega}^{\left(end\right)}:\,\left|\Phi\left\{ \underline{\Omega}^{\left(end\right)}\right\} -\Phi\left\{ \underline{\Omega}^{\left(opt\right)}\right\} \right|\leq\xi\label{eq:_SbD.goal}\end{equation}
$\Phi$ being a metric that quantifies the mismatch of a problem solution,
$\underline{\Omega}$, from the project/user requirements, $\underline{\Gamma}^{th}$,\begin{equation}
\Phi\left\{ \underline{\Omega}\right\} \triangleq\left\Vert \underline{\Gamma}^{th}-\underline{\Upsilon}\left\{ \underline{\Omega}\right\} \right\Vert ^{2}\label{Mapping Solution Space - Requirement Space}\end{equation}
where $\underline{\Upsilon}$ is the mapping function between the
solution space, $\Re\left\{ \underline{\Omega}\right\} $, and the
requirements space, $\Re\left\{ \underline{\Gamma}\right\} $ ($\underline{\Upsilon}:\,\mathcal{\Re}\left\{ \underline{\Omega}\right\} \to\Re\left\{ \underline{\Gamma}\right\} $,
being $\Re\left\{ \underline{\Gamma}\right\} \equiv\Re\left\{ \Phi\right\} $).

\noindent To find a computationally-efficient solution of such a design
problem (\ref{eq:_SbD.goal}), the \emph{SbD} exploits four interconnected
\emph{functional blocks} devoted \emph{}to the following \emph{sub-tasks}
(Fig. 1):

\begin{enumerate}
\item \noindent \emph{Requirements and Constraints Definition} (\emph{RCD})
- Mathematical definition of the project requirements, $\underline{\Gamma}^{th}$,
and of a set of physical-admissibility constraints, $G_{a}^{min}\le G_{a}\left(\underline{\Omega}\right)\le G_{a}^{max}$
($a=1,\,...,\, A$), starting from specifications, guidelines, and
objectives provided by the end-users in the statement of work (\emph{SoW});
\item \noindent \emph{Problem Formulation} (\emph{PF}) - Mathematical re-formulation
of the synthesis problem as an optimization one by (\emph{i}) selecting/defining
a parametric model of the solution, (\emph{ii}) identifying the corresponding
\emph{DoF}s, $\underline{\Omega}$, that is the minimum number of
univocally representative model descriptors, and (\emph{iii}) choosing
the cost function, $\Phi\left\{ \underline{\Omega}\right\} $, which
mathematically codes the mismatch between the project requirements/constraints
and the performance of the model, $\underline{\Upsilon}\left\{ \underline{\Omega}\right\} $,
whose global optimum corresponds to the best admissible physical solution
of the problem at hand;
\item \noindent \emph{Cost Function Computation} (\emph{CFC}) \emph{-} Evaluation,
in the most efficient and accurate way, of the cost function value
$\Phi\left\{ \underline{\Omega}^{\left(p\right)}\right\} $ of a trial
solution $\underline{\Omega}^{\left(p\right)}$ in order to {}``quantify''
the optimality of the $p$-th ($p=1,...,P$) trial physical solution
(i.e., the fitness of this solution to the problem at hand);
\item \noindent \emph{Solution Space Exploration} (\emph{SSE}) - Dealing
with an optimization problem, the task to carry out is the sampling
of the solution space, $\mathcal{\Re}\left\{ \underline{\Omega}\right\} $,
bounded by $G_{a}^{min}\le G_{a}\left(\underline{\Omega}\right)\le G_{a}^{max}$
($a=1,\,...,\, A$), to look for the global optimum of the cost function
$\Phi\left\{ \underline{\Omega}\right\} $ {[}i.e., $\underline{\Omega}^{\left(end\right)}$
in (\ref{eq:_SbD.goal}){]}. Therefore, the SSE block is aimed at
generating a succession of $I$ trial solutions \{$\underline{\Omega}_{i}^{\left(p\right)}$;
$i=1,...,I$\} ($p=1,\,...,\, P$), $P$ and $I$ being the number
of agents and of iterations, respectively, by means of a suitable
optimization algorithm so that $\underline{\Omega}^{\left(end\right)}=\arg\left\{ \min_{i=1,...,I}\left[\Phi_{i}^{best}\right]\right\} $
($\Phi_{i}^{best}\triangleq\Phi\left\{ \underline{\Omega}_{i}^{best}\right\} $)
being $\underline{\Omega}_{i}^{best}=\arg\left\{ \min_{p=1,...,P}\left[\Phi_{i}^{\left(p\right)}\right]\right\} $
($\Phi_{i}^{\left(p\right)}\triangleq\Phi\left\{ \underline{\Omega}_{i}^{\left(p\right)}\right\} $).
Generally speaking, $\underline{\Omega}_{i+1}^{\left(p\right)}=\underline{\Omega}_{i}^{\left(p\right)}+\underline{v}_{i}^{\left(p\right)}$
where $\underline{v}_{i}^{\left(p\right)}=\left\{ v_{i,\, k}^{\left(p\right)};\, k=1,\,...,\, K\right\} $
is a vectorial increment defined on the basis of a suitable set of
operators $\bm{\mathcal{L}}$, $\underline{v}_{i}^{\left(p\right)}=\bm{\mathcal{L}}\left[\left(\underline{\Omega}_{j}^{\left(q\right)},\,\Phi_{j}^{\left(q\right)}\right)\right.$;
$j=1,...,\left(i-1\right)$; $q=1,...,\left.P\right]$ depending on
the optimization strategy at hand.
\end{enumerate}

\section{\emph{SbD} Functional Blocks Implementation}

To provide the readers with a general description of the \emph{SbD}
framework and implementation strategies, let us now focus on the \emph{PF},
\emph{CFC}, and \emph{SSE} functional blocks, postponing the details
of the first block (\emph{RCD}) to the illustrative examples discussed
later on.

\subsection{Problem Formulation (\emph{PF})}

In order to synthesize satisfactory and reliable solutions, all project
requirements/constraints defined by the end-user must be carefully
{}``translated'' into a proper mathematical framework, as detailed
in the following:

\subsubsection*{- Solution Model and \emph{DoF}s Identification \label{sub:Solution-Model-and-DoF}}

In global search strategies, such as \emph{EA}s, the number $P$ of
agents/trial solutions evaluated at each $i$-th ($i=1,...,I$) iteration
is proportional to the number of \emph{DoF}s, $K$ \cite{Rocca 2009}\cite{Rocca 2011}.
Accordingly, a suitable formulation of the problem at hand must be
made so that (\emph{i}) $K$ is as low as possible to minimize the
computational burden for a more efficient design, but at the same
time, (\emph{ii}) the choice of the $K$ \emph{DoF}s, $\underline{\Omega}$,
guarantees the existence of a solution fitting the \emph{SbD} goal
(\ref{eq:_SbD.goal}) as well as its careful correspondence with a
feasible physical solution. Towards this end, parametric studies are
a valid approach to perform a \emph{sensitivity analysis} and to identify
which descriptors (varied within suitable bounds) have the highest
impact on the performance indexes, $\underline{\Upsilon}\left\{ \underline{\Omega}\right\} $,
and are the most representative to define the minimum set of $K$
descriptors. On the other hand, a key factor in the {}``representation''
of the actual physical solution is the choice of the basis functions
to be exploited for defining \emph{smarter} solution models, which
are characterized by a reduced dimensionality, $K$, while ensuring
a high flexibility in the solution representation. This is the case,
for instance, when synthesizing the shape/profile of an \emph{EM}
device as in \cite{Salucci 2019} where linear arrays have been miniaturized
by means of isotropic covering meta-lenses. More in detail, the goal
has been that of determining the optimal transformation-optics 2\emph{D}
profile, $\gamma\left(x,\, y\right)$, such that the lens-enclosed
antenna mimic a reference one with larger aperture. By describing
$\gamma\left(x,\, y\right)$ with a \emph{pixel-basis} representation\begin{equation}
\gamma^{\left(pix\right)}\left(x,\, y\right)=\sum_{k=1}^{K^{\left(pix\right)}}\Omega_{k}^{\left(pix\right)}B_{k}^{\left(pix\right)}\left(x,\, y\right)\label{eq: Pixel Representation}\end{equation}
where $B_{k}^{\left(pix\right)}\left(x,\, y\right)=1$ if $\left(x,\, y\right)\in\rho_{k}$,
while $B_{k}^{\left(pix\right)}\left(x,\, y\right)=0$ otherwise,
$\rho_{k}$ being the $k$-th discretization cell of the lens, the
arising number of \emph{DoF}s was equal to the number of pixels, $K^{\left(pix\right)}$
{[}Fig. 2(\emph{a}){]}. Differently, a \emph{spline-based} representation
has been adopted in \cite{Salucci 2019} to describe the lens profile
in terms of Bezier quadratic curves by means of second order polynomials,
$B_{k}^{\left(spl\right)}\left(x,\, y,\,\underline{\Omega}^{\left(spl\right)}\right)$
($k=1,\,...,\, K^{\left(spl\right)}$) \cite{Salucci 2018}\cite{Lizzi 2008}\cite{Massa 2018}
so that the \emph{DoF}s coincide with the \emph{control points} of
the spline curve, $\underline{\Omega}^{\left(spl\right)}=\left\{ \left(x_{k}^{\left(spl\right)},\, y_{k}^{\left(spl\right)}\right);\, k=1,\,...,\, K^{\left(spl\right)}\right\} $
{[}Fig. 2(\emph{b}){]}. It is worth pointing out that while the pixel-based
representation (\ref{eq: Pixel Representation}) needs a significantly
larger number of descriptors, $K^{\left(pix\right)}\gg K^{\left(spl\right)}$,
to yield a detailed model of $\gamma\left(x,\, y\right)$, the use
of spline bases allows one to model a continuous profile with a limited
number of control points (e.g., $K^{\left(spl\right)}=5$ \cite{Salucci 2019}).
Of course, using a spline-based representation is not always the best
solution, but certainly a careful study on the representation properties
of a set of basis functions is a {}``\emph{golden}'' rule for having
a competitive \emph{SbD}-based design approach;

\subsubsection*{- Cost Function Definition \label{sub:Cost-Function-Definition}}

The cost function $\Phi\left\{ \underline{\Omega}\right\} $ is the
unique link between the optimization strategy and the physics of the
\emph{EM} problem. Therefore, it must be carefully defined to guarantee
the correct sampling of the solution space and the achievement of
feasible solutions \cite{Rocca 2009}. Moreover, the choice of the
cost function determines the overall \emph{complexity} of the $K$-dimensional
\emph{landscape} explored by the \emph{SSE} block. To the best of
authors' knowledge, there is not a general guideline to optimally
select $\Phi\left\{ \underline{\Omega}\right\} $ as well as to have
the arising cost function with suitable properties, for instance,
to limit the occurrence of local minima/false solutions. However,
similarly to what is done in inverse scattering \cite{Isernia 2001}\cite{Salucci 2017b},
its behavior can be roughly \emph{estimated} by analyzing the functional
cuts along some directions of the solution space\begin{equation}
\Phi\left\{ t,\,\underline{\Omega}^{\left(1\right)},\,\underline{\Omega}^{\left(2\right)}\right\} =\Phi\left\{ \left(1-t\right)\times\underline{\Omega}^{\left(1\right)}+t\times\underline{\Omega}^{\left(2\right)}\right\} \label{eq:functional-cut}\end{equation}
$\underline{\Omega}^{\left(1\right)}$ and $\underline{\Omega}^{\left(2\right)}$
being two user-chosen positions within the solution space, while $t$
is a real variable. From (\ref{eq:functional-cut}), it turns out
that $\Phi\left\{ t,\,\underline{\Omega}^{\left(1\right)},\,\underline{\Omega}^{\left(2\right)}\right\} =\Phi\left\{ \underline{\Omega}^{\left(1\right)}\right\} $
if $t=0$ and $\Phi\left\{ t,\,\underline{\Omega}^{\left(1\right)},\,\underline{\Omega}^{\left(2\right)}\right\} =\Phi\left\{ \underline{\Omega}^{\left(2\right)}\right\} $
if $t=1$, while sweeping $t$ within suitable bounds gives some insights
on the behavior of the cost function along a one-dimensional (1\emph{-D})
cut passing through $\underline{\Omega}^{\left(1\right)}$ and $\underline{\Omega}^{\left(2\right)}$.
What is the reason for doing it? On the one hand, such evaluations
allows one to have some indications (not analytic proofs) on the degree
of complexity/nonlinearity of the functional space at hand %
\footnote{\noindent The \emph{feasibility} of such analyses clearly depends
on the computational complexity and \emph{CPU}-time of each evaluation
of the cost function. Of course, careful \emph{a-priori} analyses
must be performed case-by-case to infer about their worthiness and
proper set-up. %
}. On the other hand, they can provide a valid support for the optimal
choice and implementation of the \emph{CFC} and \emph{SSE} blocks.

\subsection{Cost Function Computation (\emph{CFC}) \label{sub:Cost-Function-Computation}}

The computation of $\Phi\left\{ \underline{\Omega}\right\} $ requires
the evaluation of specific performance indexes of the synthesized
\emph{EM} device. Towards this aim, several numerical techniques are
available for performing accurate \emph{FW} analyses \cite{Jin 2014}\cite{Harrington 1993}\cite{Taflove 2005}.
By formulating forward \emph{EM} problems by means of properly discretized
sets of integral and/or differential equations, the corresponding
numerical solution needs the computation of thousands or millions
of unknowns. Therefore, a repeated evaluation of the cost function
$\Phi\left\{ \underline{\Omega}\right\} $ is the real \emph{bottleneck}
of standard (\emph{StD}) optimization-based designs whose computational
burden is\begin{equation}
\Delta t_{StD}=\left(P\times I\right)\times\Delta t_{FW}\label{eq:}\end{equation}
$\Delta t_{FW}$ being the \emph{CPU} time for a single \emph{FW}
simulation. Unless reliable analytic model are available, learning
by examples (\emph{LBE}) techniques are exploited by the \emph{SbD}
to significantly reduce the computational burden, while keeping a
reliable prediction of the performance of the synthesized device/system.
In short, \emph{LBE}s are devoted to build fast \emph{surrogate models}
(\emph{SM}s) able to \emph{predict}, in a computationally-efficient
fashion, the outcome of high-fidelity \emph{EM} simulations \cite{Massa 2018b}.
From an architectural viewpoint, \emph{LBE} strategies are two-step
implementations composed by (\emph{i}) a training and (\emph{ii})
a testing phase. The training phase is typically performed \emph{off-line}
and it is devoted to build an accurate and fast \emph{surrogate} of
the cost function $\Phi\left\{ \underline{\Omega}\right\} $, $\widetilde{\Phi}\left\{ \underline{\Omega}\right\} $,
starting from a training set of $S$ examples/observations of the
input/output (\emph{I/O}) relationship, $\mathcal{D}_{S}=\left[\left(\underline{\Omega}^{\left(s\right)};\,\Phi^{\left(s\right)}\right);\,\,\, s=1,\,...,\, S\right]$
where $\Phi^{\left(s\right)}$ stands for $\Phi^{\left(s\right)}\triangleq\Phi\left\{ \underline{\Omega}^{\left(s\right)}\right\} $
(Fig. 3). During the test phase, \emph{on-line} predictions of the
cost function value are then outputted for previously-unseen inputs
\cite{Massa 2018b}. 

\noindent Among several \emph{LBE} strategies, let us focus in the
following on the most commonly-adopted ones in \emph{EM} engineering
\cite{Massa 2018b}. Radial Basis Function Networks (\emph{RBFN}s)
are popular artificial neural networks (\emph{ANN}s) computing the
surrogate $\widetilde{\Phi}\left\{ \underline{\Omega}\right\} $ as
a linear combination, through suitable real expansion coefficients,
\{$w^{\left(s\right)}$; $s=1,...,S$\}, of $S$ Gaussian functions
\cite{Massa 2018b}, \{$\psi^{\left(s\right)}\left\{ \underline{\Omega}\right\} $;
$s=1,...,S$\},\begin{equation}
\widetilde{\Phi}\left\{ \underline{\Omega}\right\} =\sum_{s=1}^{S}\psi^{\left(s\right)}\left\{ \underline{\Omega}\right\} w^{\left(s\right)}.\label{eq:RBFN}\end{equation}
Otherwise, Support Vector Regressors (\emph{SVR}s) define the surrogate
model as follows\begin{equation}
\widetilde{\Phi}\left\{ \underline{\Omega}\right\} =\sum_{s=1}^{S}\left[\left(\alpha^{\left(s\right)}-\beta^{\left(s\right)}\right)\mathcal{K}\left\{ \underline{\Omega}^{\left(s\right)},\,\underline{\Omega}\right\} \right]+\varsigma\label{eq:SVR}\end{equation}
where $\alpha^{\left(s\right)}$ and $\beta^{\left(s\right)}$($s=1,...,S$)
are the \emph{SVR} weights, while $\mathcal{K}\left\{ \underline{\Omega}^{\left(s\right)},\,\underline{\Omega}\right\} $
is the kernel function, $\varsigma$ being a bias \cite{Massa 2018b}.
A main difference between \emph{RBFN}s and \emph{SVR}s is the intrinsic
capability of \emph{RBFN}s to exactly fit/interpolate the training
samples (i.e., $\widetilde{\Phi}\left\{ \underline{\Omega}^{\left(s\right)}\right\} =\Phi\left\{ \underline{\Omega}^{\left(s\right)}\right\} $,
$s=1,...,S$). Otherwise, the \emph{SVR} tolerates/neglects deviations
of the surrogate prediction $\widetilde{\Phi}\left\{ \underline{\Omega}\right\} $
from the actual cost function $\Phi\left\{ \underline{\Omega}\right\} $
smaller than a threshold $\epsilon$ by defining an {}``$\epsilon$-insensitive
tube'' \cite{Massa 2018b}. 

\noindent Of course, there is not an optimal and unique choice for
the best prediction technique, but this depends on the design problem
at hand as well as on the selection of the remaining \emph{SbD} blocks.
Indeed, exactly performing like high-fidelity \emph{FW} simulators
when processing previously-explored solutions may be a desirable feature
since the \emph{I/O} relationship, $\underline{\Upsilon}\left\{ \underline{\Omega}\right\} $,
is purely deterministic. However, the \emph{SM} should not be regarded
in the \emph{SbD} framework as a highly-reliable computationally-efficient
alternative to \emph{FW} solvers, but rather as a sufficiently-accurate
estimator of the behavior of the cost function to guide the solution-space
sampling/exploration towards the \emph{attraction basin} of $\underline{\Omega}^{\left(opt\right)}$.
In order to better understand this latter concept, let us consider
some simple yet intuitive examples on well-known 1\emph{D} ($K=1$)
benchmark cost functions, $\Phi\left\{ \underline{\Omega}\right\} \triangleq\Phi\left\{ \Omega_{1}\right\} $.
Figure 4(\emph{a}) shows the 1\emph{D} Levy's cost function within
the range $\Omega_{1}\in\left[-10,\,10\right]$ \cite{Laguna 2005}
along with the predictions made by the \emph{RBFN} and the \emph{SVR}
surrogates starting from $S=6$ randomly-chosen training samples.
As it can be observed, the \emph{SVR} correctly identifies the presence
of a \emph{valley} centered at the global minimum of the actual cost
function, $\Omega_{Levy}^{\left(opt\right)}=1$, while a significantly
worse prediction of the cost function behavior is given by the \emph{RBFN}
even though this latter perfectly fits all training observations.
However, the \emph{SVR} may lead to an over-smoothed surrogate of
$\Phi$ failing to {}``understand'' the overall trend of the actual
cost function as shown in Fig. 4(\emph{b}) for the Schwefel's function
\cite{Laguna 2005} ($\Omega\in\left[-500,\,500\right]$, $\Omega_{Schwefel}^{\left(opt\right)}=420.9687$).
The two surrogates perform similarly when dealing with the Ackley's
function \cite{Laguna 2005} {[}$\Omega\in\left[-5,\,5\right]$, $\Omega_{Ackley}^{\left(opt\right)}=0$
- Fig. 4 (\emph{c}){]}.

\noindent Another widely-used \emph{LBE} method is the Ordinary Kriging
(\emph{OK}) whose remarkable advantage over the \emph{RBFN} and the
\emph{SVR} is the straightforward capability of providing a measure,
$\Psi\left\{ \underline{\Omega}\right\} $, of the degree of \emph{reliability/confidence}
associated to any prediction $\widetilde{\Phi}\left\{ \underline{\Omega}\right\} $
\cite{Massa 2018b}\cite{Jones 1998}. As it will be explained in
the next Sections, such an additional output is a powerful source
of information to be profitably exploited to enhance the effectiveness
of the whole \emph{SbD} synthesis. More in detail, the surrogate model
generated by the \emph{OK} is given by \cite{Forrester 2008}\cite{Jones 1998}\begin{equation}
\widetilde{\Phi}\left\{ \underline{\Omega}\right\} =\mu+\underline{\eta}^{T}\underline{\underline{\mathcal{R}}}^{-1}\left(\underline{\Phi}-\underline{\mathcal{I}}\mu\right)\label{eq:OK}\end{equation}
where $\mu$ is a real constant, $\underline{\Phi}=\left[\Phi^{\left(s\right)};\, s=1,...,S\right]^{T}$,
$.{}^{T}$ being the transpose operator, $\underline{\mathcal{I}}$
is the $\left(S\times1\right)$ unit vector, $\underline{\underline{\mathcal{R}}}=\left\{ \mathcal{R}_{pq};\, p,\, q=1,\,...,\, S\right\} $
is the $\left(S\times S\right)$ training correlation matrix, and
$\underline{\eta}=\left\{ \eta_{s};\, s=1,...,S\right\} $ is the
correlation vector of $\underline{\Omega}$ \cite{Jones 1998}. As
for the prediction reliability metric, its meaning is quite intuitive
since it is defined as the weighted distance between $\underline{\Omega}$
and the $S$ training samples (i.e., $\Psi\left\{ \underline{\Omega}\right\} \propto\frac{\sum_{s=1}^{S}\left\Vert \underline{\Omega}-\underline{\Omega}^{\left(s\right)}\right\Vert _{2}}{S}$)
being $\Psi\left\{ \underline{\Omega}\right\} =0$ and $\widetilde{\Phi}\left\{ \underline{\Omega}\right\} =\Phi\left\{ \underline{\Omega}\right\} $
only if $\underline{\Omega}=\underline{\Omega}^{\left(s\right)}$
($s=1,...,S$) (Fig. 5). More specifically, the \emph{OK} uncertainty
is modeled starting from the assumption that the cost function value
$\Phi\left\{ \underline{\Omega}\right\} $ is the realization of a
normally-distributed random variable with mean $\widetilde{\Phi}\left\{ \underline{\Omega}\right\} $
and standard deviation equal to $\Psi\left\{ \underline{\Omega}\right\} $
\cite{Forrester 2008}\cite{Jones 1998} (Fig. 5). For the sake of
completeness, the predictions made by the \emph{OK} for the Levy's,
Schwefel's, and Ackley's functions are reported in Fig. 4.

\subsubsection*{- \emph{SbD}-Driven Training Set Generation \label{sub:Sampling-Strategies}}

In order to yield a reliable \emph{SM,} the training set must be properly
built. Hypothetically, only a very large (ideally infinite) number,
$S$, of \emph{I/O} pairs would allow to exactly predict the cost
function value $\Phi\left\{ \underline{\Omega}\right\} $ (i.e., $\widetilde{\Phi}\left\{ \underline{\Omega}\right\} \to\Phi\left\{ \underline{\Omega}\right\} $
when $S\to\infty$). Practically, strategies for the selection of
the minimum number of \emph{representative} samples guaranteeing the
\emph{SM} prediction error is below a desired threshold are needed.
Towards this aim, the \emph{SbD} toolkit deals with \emph{LBE} strategies
from a different perspective and in terms of a \emph{three-step} approach
where only the last one is finalized at defining the \emph{SM} \cite{Salucci 2016}.
More in detail, the first step is concerned with the reduction of
the dimension of the input (solution) space $\mathcal{\Re}\left\{ \underline{\Omega}\right\} $
to mitigate the negative effects of the \emph{curse of dimensionality}
\cite{Rodriguez 2007}\emph{.} It is worth pointing out that such
a task is partially/preliminarily addressed by the \emph{PF} block
when choosing the smallest set of the most representative \emph{DoF}s
that guarantees the existence of a physically-admissible solution
(Sect. \ref{sub:Solution-Model-and-DoF}). However, whether $K$ is
still high (e.g., $K>20$) lower-cardinality yet highly-informative
training sets can be built by means of \emph{space reduction} techniques.
These latter determine a reduced set of $K'$ ($K'\ll K$) \emph{DoF}s,
called \emph{reduced features}, \emph{}$\underline{\chi}^{\left(s\right)}=\left\{ \chi_{k}^{\left(s\right)};\, k=1,...,K'\right\} $
by means of a linear/non-linear transformation operator $\Lambda$
so that a \emph{reduced} database, $\widehat{\mathcal{D}}_{S}=\left\{ \left(\underline{\chi}^{\left(s\right)},\,\Phi^{\left(s\right)}\right);\, s=1,...,S\right\} $
(Fig. 3), is built. The $\Lambda$-based mapping can be \emph{function-independent}
(i.e., $\underline{\chi}^{\left(s\right)}=\Lambda\left\{ \underline{\Omega}^{\left(s\right)}\right\} $),
as in the Principal Component Analysis (\emph{PCA}) \cite{Salucci 2016}
and in the Sammon Mapping (\emph{SAM}) \cite{Liu 2014}, or \emph{function-dependent}
(i.e., $\underline{\chi}^{\left(s\right)}=\Lambda\left\{ \underline{\Omega}^{\left(s\right)},\,\Phi^{\left(s\right)}\right\} $)
as for the Partial Least Squares (\emph{PLS}) \cite{Salucci 2016}.
\emph{}The second step is aimed at an {}``exhaustive'' representation
of the input space $\mathcal{\Re}\left\{ \underline{\Omega}\right\} $
by properly selecting the $S$ \emph{I/O} pairs. The available sampling
strategies can be classified into two main categories: (\emph{i})
one-shot/non-iterative and (\emph{ii}) adaptive strategies \cite{Lam 2008}\cite{Crombecq 2011b}\cite{Garud 2017}\cite{Liu 2017}\cite{Maljovec 2013}.
The uniform grid sampling belongs to the first class (\emph{i}) and
it performs a \emph{full-factorial} exploration of the input space
by uniformly sampling each $k$-th ($k=1,...,K'$) dimension and considering
all existing combinations. Clearly, it becomes rapidly \emph{unfeasible}
when increasing the solution space dimensionality $K'$ and/or the
number of quantization levels, $Q$ (e.g., $S=Q^{K'}=10^{5}$ when
$K'=5$ and $Q=10$). To overcome such a drawback, other strategies
such as the Latin Hypercube Sampling (\emph{LHS}) \cite{Liu 2017}
and the Orthogonal Arrays (\emph{OA}s) method \cite{Bayraktar 2011}\cite{Salucci 2019b}
can be exploited. These approaches, which belong to the class (\emph{ii}),
are based on the iterative selection of the $S$ training samples
to reach a profitable balancing between \emph{exploration} (i.e.,
new samples in the regions of $\mathcal{\Re}\left\{ \underline{\chi}\right\} $
where the sampling rate is low) and \emph{exploitation} {[}i.e., new
samples where the cost function $\Phi$ is more nonlinear as it can
be inferred from the cuts-analysis (\ref{eq:functional-cut}){]} \cite{Liu 2017};

\subsubsection*{- \emph{SbD} Time Saving}

\noindent When using a \emph{SM} instead of the \emph{FW} solver,
the total cost of the \emph{SbD}-based synthesis turns out to be\begin{equation}
\Delta t_{SbD}=\Delta t_{SM}\left(S\right)+\left(P\times I\right)\times\Delta t_{test}\label{eq: SbD Time}\end{equation}
where $\Delta t_{test}$ is the \emph{CPU} time to yield a single
$\Phi$-prediction and $\Delta t_{SM}\left(S\right)=\left(S\times\Delta t_{FW}\right)+\Delta t_{train}$
is the \emph{CPU} time to perform the $S$ training simulations and
to generate the \emph{SM} %
\footnote{\noindent It should be noticed that also the training/test times are
functions of $S$ {[}i.e., $\Delta t_{train/test}=\Delta t_{train/test}\left(S\right)${]}.
However, $S$ is kept quite small ($S\ll10^{3}$) in practical applications
to guarantee a significant time saving. Therefore, the dependence
on the training set size $S$ can be neglected since $\Delta t_{train/test}\left(S\right)\ll\left(S\times\Delta t_{FW}\right)$.%
}. Accordingly, the \emph{SbD} becomes profitable and very competitive
when $\Delta t_{SbD}\ll\Delta t_{StD}$ being\begin{equation}
\Delta t_{StD}=\left(P\times I\right)\times\Delta t_{FW}.\label{eq: StD Time}\end{equation}
To provide a simple, although rigorous, indication of the overall
time saving of the \emph{SbD} with respect to a trivial integration
of a \emph{FW} solver within an optimization loop, let us consider
that in practical situations $\left(\Delta t_{train},\,\Delta t_{test}\right)\ll\Delta t_{FW}$,
thus the following approximation generally holds true: $\Delta t_{SbD}\approx S\times\Delta t_{FW}$.
Accordingly, the time saving percentage thanks to the \emph{SbD} (i.e.,
$\Delta t_{sav}\triangleq\frac{\Delta t_{StD}-\Delta t_{SbD}\left(S\right)}{\Delta t_{StD}}$)
is equal to\begin{equation}
\Delta t_{sav}\approx\frac{\left(P\times I\right)-S}{\left(P\times I\right)}\times100,\label{eq: SbD Time saving}\end{equation}
thus the rule-of-thumb applicability condition for the \emph{SbD}
is $S<\left(P\times I\right)$.

\noindent The percentage saving in (\ref{eq: SbD Time saving}) can
be even higher in case of massive parallel computing. With reference
to the computational scenario where $O$ ($O\ge P$) processors are
available, while the effects of multiple and parallel computational
capabilities can be exploited by a standard approach only to reduce
the optimization time\begin{equation}
\Delta t_{StD}^{\parallel}=\frac{\Delta t_{StD}}{P}\label{eq: || StD Time}\end{equation}
by sharing among the $O$ \emph{CPU}s the evaluation of the cost function
of the $P$ multiple-agents/trial-solutions at each $i$-th ($i=1,...,I$)
iteration%
\footnote{\noindent Indeed, the number of \emph{CPU}s that can be used in parallel
in standard optimization algorithms is bounded by the number of trial
solutions $P$ since the generation of new solutions depends on the
outcomes of the previous iteration.%
}, the \emph{SbD} benefits of this computational over-boost also for
reducing the training time\begin{equation}
\Delta t_{SbD}^{\parallel}=\frac{\Delta t_{SM}\left(S\right)}{O}+\frac{\left(P\times I\right)\times\Delta t_{test}}{P}.\label{eq: || SbD Time}\end{equation}
Therefore, the percentage time saving is roughly around

\noindent \begin{equation}
\Delta t_{sav}^{\parallel}\approx\left(1-\frac{S}{O\times I}\right)\times100\label{eq: || SbD Time saving}\end{equation}
{[}i.e., $\Delta t_{sav}^{\parallel}>0$ if $S<\left(O\times I\right)${]}
and it further reduces to $\Delta t_{sav}^{\parallel}\approx\left(1-\frac{1}{I}\right)\times100$
{[}i.e., $\Delta t_{sav}^{\parallel}>0$ always since $I>1$ by definition
of iterative optimization loop{]}, if $O\ge S$ since $\Delta t_{SM}^{\parallel}\left(S\right)\approx\Delta t_{FW}$.

\subsection{Solution Space Exploration (\emph{SSE}) \label{sub:Solution-Space-Exploration}}

Many global optimization methods exist and the choice of the most
effective algorithm for the synthesis problem under study is a key
issue and not trivial task at all. From a theoretical viewpoint, the
{}``\emph{no free-lunch theorem}'' (\emph{NFL}) of optimization
\cite{Wolpert 1997} states that (\emph{i}) {}``\emph{the average
performance of any pair of algorithms across all problems is identica}l''
and (\emph{ii}) {}``\emph{no matter what the cost function, by simply
observing how well the algorithm has done so far tells us nothing
about how well it would do if we continue to use it on the same cost
function}''. This implies that {}``\emph{whether an optimization
algorithm performs better than random search on some class of problems,
then it must perform worse than random search on another class}''
\cite{Wolpert 1997}. Therefore, the application of an arbitrary algorithm
to an optimization problem without understanding the nature and the
properties/features of the cost function $\Phi$ as well as of $\underline{\Omega}$
is on average equivalent to perform a random search. Indeed, it is
well proven that each optimizer has its own {}``optimal niche''
of application where it outperforms other alternatives or vice-versa
a suitable reformulation of the synthesis problem at hand allows one
to use a particular optimization tool. By extension, the \emph{NFL}
principles hold true for the \emph{SbD}, as well, since this latter
formulates a synthesis problem as an optimization one. More specifically,
the \emph{NFL} rules apply to the SbD framework as follows {}``\emph{it
is not possible to a-priori identify the best combination of the functional
blocks of the} SbD \emph{able to perform well on every possible problem}''.

\noindent To give some insights on this concept and its consequences,
let us focus our attention to the integration of two representative
blocks, namely the \emph{SM} and the optimization tool. Towards this
end, let us consider the optimization of the three benchmark functions
in Fig. 4, but with $K=6$, performed with the {}``bare'' integration
of the \emph{RBFN}, the \emph{SVR}, and the \emph{OK} models with
two state-of-the-art evolutionary optimizers, namely the Particle
Swarm Optimizer (\emph{PSO}) and the Differential Evolution (\emph{DE})
\cite{Rocca 2009}\cite{Rocca 2011}. Given the stochastic nature
of both the \emph{SM}s (i.e., the \emph{LHS}) and the optimizers,
the median realization over $R=20$ executions with $P=10$ agents
for $I=200$ iterations is reported to yield statistically-meaningful
results. Figure 6 reports the cost function values at the last iteration
($i=I$) of the different \emph{BARE-SbD} algorithms versus the $\frac{S}{K}$
ratio. Dealing with the Levy's function {[}Fig. 6(\emph{a}){]}, the
choice of the \emph{SVR} to generate the \emph{SM} turns out to be
successful for the arising \emph{SbD} implementation \emph{}since
both the \emph{PSO-SVR} and the \emph{DE-SVR} integrations yield the
best results provided that a sufficient number of training samples
is available (i.e., $\frac{S}{K}\geq20$ $\rightarrow$ $S\geq120$
being $K=6$). Otherwise, the \emph{SVR}-based methods are the worse
ones for the minimization of the Schwefel's function {[}Fig. 6(\emph{b}){]}
because of the \emph{over-smoothing} in approximating the actual cost
function {[}Fig. 4(\emph{b}){]}. On the contrary, the \emph{OK} enables
a proper exploration of the solution space when integrated with the
\emph{PSO} {[}Fig. 6(\emph{b}){]}. This latter choice is sub-optimal
when dealing with the Ackley's function, while both \emph{SVR}-based
strategies perform very well whatever the cardinality $S$ of the
training set {[}Fig. 6(\emph{c}){]}.

\section{Advanced \emph{SbD} Strategies \label{sec:Advanced-SbD-Strategies}}

Totally replacing the \emph{FW} solver with a \emph{SM}, as done in
\emph{BARE-SbD} approaches (Sect.\ref{sub:Solution-Space-Exploration}),
may lead to sub-optimal designs \cite{Jin 2011}. This holds true
especially when non-negligible time savings are required to permit
the synthesis of high-complexity and computationally demanding devices/systems,
thus setting a low size, $S$, for the training set because of the
very limited number of \emph{affordable} simulations within a reasonable
amount of time. As a matter of fact, treating the \emph{SM} as a {}``magic
black box'' could produce undesired effects such as (\emph{i}) the
convergence towards false solutions/local minima and/or (\emph{ii})
the prediction of \emph{unfeasible} cost function values (e.g., negative
values, $\widetilde{\Phi}\left\{ \underline{\Omega}\right\} <0$).
Moreover, simply increasing $S$ to \emph{globally} improve the prediction
accuracy could be not enough to prevent such issues and advanced \emph{SbD}
strategies are necessary. They are mainly based on the following basic
\emph{}recipes: \emph{}(\emph{i}) the \emph{local refinement} of the
\emph{SM} within the \emph{attraction basin} of $\underline{\Omega}^{\left(opt\right)}$
and/or (\emph{ii}) the \emph{interactive} \emph{collaboration} between
the optimizer and the predictor. By following those guidelines, two
advanced \emph{SbD} implementations are described in the following.

\subsection{Optimization-Driven {}``Smart'' \emph{SM} Generation}

Unlike state-of-the-art adaptive sampling strategies (Sect. \ref{sub:Sampling-Strategies}),
the accuracy of the \emph{SM} is \emph{locally} enhanced in the neighborhood
of $\underline{\Omega}^{\left(opt\right)}$. Towards this end, an
\emph{optimization-driven} adaptive sampling strategy can be adopted
to build the training set $\mathcal{D}_{S}$ by iteratively adding
new samples to an initial dataset $\mathcal{D}_{0}$ with $S_{0}$
($S_{0}<S$) \emph{I/O} pairs. Such a strategy belongs to the class
of the {}``output space filling'' (\emph{OSF}) techniques \cite{Bilicz 2010}
and it is aimed at uniformly exploring the output/cost function space
for which $\Phi\left\{ \underline{\Omega}\right\} <\Phi_{th}$, $\Phi_{th}$
being a user-specified threshold. It is performed within the \emph{CFC}
block (Fig. 1) and it consists of the following procedural steps:

\begin{enumerate}
\item \emph{Initialization} - Generate $\mathcal{D}_{0}=\left[\left(\underline{\Omega}^{\left(s\right)};\,\Phi^{\left(s\right)}\right);\, s=1,...,S_{0}\right]$
by sampling the input space via \emph{LHS} and initialize the loop
index ($j=1$);
\item \emph{SbD-OSF Loop} {[}$j=1,...,\left(S-S_{0}\right)${]}

\begin{enumerate}
\item Train a \emph{SM} using the $S_{j-1}$ samples of the dataset $\mathcal{D}_{j-1}$; 
\item Sample the input space via \emph{LHS} to generate $C$ \emph{candidates},
\{$\underline{\Omega}^{\left(c\right)}$; $c=1,...,C$\}%
\footnote{According to the reference literature \cite{Maljovec 2013}, the number
of candidates is set to a value in the range $C\in\left[50,\,200\right]\times K$.%
}, and predict the corresponding cost function values, \{$\widetilde{\Phi}^{\left(c\right)}\triangleq\widetilde{\Phi}\left\{ \underline{\Omega}^{\left(c\right)}\right\} $;
$c=1,...,C$\};
\item Select the {}``best'' candidate $\underline{\Omega}^{*}$ as $\underline{\Omega}^{*}=\arg\left\{ \max_{c=1,\,...,\, C}\left[\min_{s=1,\,...,\, S_{j-1}}\left(d_{c,\, s}\right)\right]\right\} $
subject to $\widetilde{\Phi}^{\left(c\right)}\leq\Phi_{th}$ being
$d_{c,\, s}\triangleq\left|\widetilde{\Phi}^{\left(c\right)}-\Phi^{\left(s\right)}\right|$;
\item Compute the actual cost function value of $\underline{\Omega}^{*}$,
$\Phi^{*}\triangleq\Phi\left\{ \underline{\Omega}^{*}\right\} $,
and update the training size, $S_{j}\leftarrow\left(S_{j-1}+1\right)$,
and dataset, $\mathcal{D}_{j}\leftarrow\mathcal{D}_{j-1}\bigcup\left(\underline{\Omega}^{*},\,\Phi^{*}\right)$
along with the loop index, $j\leftarrow\left(j+1\right)$. Stop the
procedure if $S_{j}=S$ or go to Step 2(a). 
\end{enumerate}
\end{enumerate}
Figure 7 illustrates the results of such a procedure when applied
to the 1\emph{-D} Ackley's function. More specifically, the \emph{SbD-OSF}
has been run by setting $S_{0}=5$ and $S=50$, while the \emph{OK}-based
\emph{SM} model has been chosen to predict $C=200$ candidates at
each $j$-th {[}$j=1,...,\left(S-S_{0}\right)${]} loop. As it can
be observed, there is an adaptive exploration of the region where
$\Phi\left\{ \underline{\Omega}\right\} <\Phi_{th}$ {[}$\Phi_{th}=6.0$
- Fig. 7(\emph{a}){]} and, as a by-product, the accuracy of the \emph{OK}
has been enhanced only close to $\underline{\Omega}^{\left(opt\right)}$
{[}Fig. 7(\emph{b}){]}.

\subsection{{}``Confidence-Enhanced'' \emph{SbD} Optimization}

Another strategy to improve the effectiveness of the \emph{SbD} is
to implement a more \emph{interactive} framework by enforcing a \emph{bilateral}
exchange of information between the \emph{CFC} and the \emph{SSE}
blocks (Fig. 1). Let us first notice that, on the one hand, the \emph{SM}
is a computationally-cheap alternative to the \emph{FW} solver, on
the other hand, the optimizer progressively localizes {}``\emph{promising}''
regions of the search space (i.e., the attraction basins of the landscape
of the cost function) where the prediction accuracy should be enhanced
to facilitate the convergence towards the global minimum, $\underline{\Omega}^{\left(opt\right)}$.
In order to fully exploit these features, a novel \emph{SbD} strategy
is proposed hereinafter by profitably combining the global search/hill-climbing
features of the \emph{PSO} \cite{Rocca 2009} with the capability
of the \emph{OK} to provide a \emph{reliability} index (Sect. \ref{sub:Cost-Function-Computation})
of the \emph{SM} predictions. The arising {}``confidence-enhanced''
\emph{PSO-OK} (\emph{PSO-OK}/\emph{C}) method is based on a {}``reinforced
learning'' (\emph{RL}) strategy that updates the \emph{SM} during
the optimization by adaptively selecting trial solutions to be evaluated
with the \emph{FW} solver. It works as follows:

\begin{enumerate}
\item \emph{Initialization} ($i=0$) - Train an \emph{OK}-based \emph{SM}
starting from an initial training set of $S_{0}$ \emph{I/O} pairs,
$\mathcal{D}_{S_{0}}$. Compute the best cost function value of the
solutions in $\mathcal{D}_{S_{0}},$$\Phi_{best}^{train}=\min_{s=1,...,S_{0}}\left[\Phi^{\left(s\right)}\right]$.
Given the desired time saving $\Delta t_{sav}$, set the maximum number
of \emph{affordable} simulations, $S$, accordingly {[}i.e., $S=P\times I\times\left(1-\frac{\Delta t_{sav}}{100}\right)${]}.
Define an initial swarm of $P$ particles with random positions \{$\underline{\Omega}_{0}^{\left(p\right)}$;
$p=1,...,P$\} and velocities \{$\underline{v}_{0}^{\left(p\right)}$;
$p=1,...,P$\};
\item \emph{PSO-OK/C Optimization Loop} ($i=0,...,I$)

\begin{enumerate}
\item For each $p$-th ($p=1,...,P$) particle, predict the cost function
value, $\widetilde{\Phi}_{i}^{\left(p\right)}$ ($\widetilde{\Phi}_{i}^{\left(p\right)}\triangleq\widetilde{\Phi}\left\{ \underline{\Omega}_{i}^{\left(p\right)}\right\} $),
and compute the associated \emph{confidence} index, $\Psi_{i}^{\left(p\right)}$
($\Psi_{i}^{\left(p\right)}\triangleq\Psi\left\{ \underline{\Omega}_{i}^{\left(p\right)}\right\} $).
If $S_{i}<S$ then go to Step 2(b), otherwise go to Step 2(c);
\item Select the {}``most promising'' particle as $\underline{\Omega}_{i}^{*}=\arg\left[\min_{p=1,\,...,\, P}\left(\mathcal{F}^{-}\left\{ \underline{\Omega}_{i}^{\left(p\right)}\right\} \right)\right]$,
where $\mathcal{F}^{-}\left\{ \underline{\Omega}_{i}^{\left(p\right)}\right\} =\widetilde{\Phi}_{i}^{\left(p\right)}-\zeta\Psi_{i}^{\left(p\right)}$
is the {}``lower confidence bound{}`` (\emph{LCB}) of the $p$-th
($p=1,...,P$) trial solution, $1\leq\zeta\leq3$ being a real coefficient
\cite{Liu 2014}. If $\mathcal{F}^{-}\left\{ \underline{\Omega}_{i}^{\left(p\right)}\right\} >\Phi_{best}^{train}$
then jump to Step 2(\emph{c}), otherwise perform the following \emph{RL}
operations:

\begin{enumerate}
\item Use the \emph{FW} solver to compute the actual cost function of $\underline{\Omega}_{i}^{*}$,
$\Phi_{i}^{*}$ ($\Phi_{i}^{*}\triangleq\Phi\left\{ \underline{\Omega}_{i}^{*}\right\} $);
\item Update the training set, $\mathcal{D}_{S_{i}}\leftarrow\mathcal{D}_{S_{i-1}}\bigcup\left(\underline{\Omega}^{*},\,\Phi_{i}^{*}\right)$,
and its size, $S_{i}\leftarrow\left(S_{i-1}+1\right)$. If $\Phi_{i}^{*}<\Phi_{best}^{train}$,
then update $\Phi_{best}^{train}$ ($\Phi_{best}^{train}\leftarrow\Phi_{i}^{*}$);
\item Re-train the \emph{OK} model with $\mathcal{D}_{S_{i}}$;
\end{enumerate}
\item Update the personal best position of each $p$-th ($p=1,...,P$) particle,
$\underline{b}_{i}^{\left(p\right)}$ ($\underline{b}_{i}^{\left(p\right)}=\arg\left\{ \min_{j=1,...,i}\left[\Phi_{j}^{\left(p\right)}\right]\right\} $),
according to the rules summarized in Tab. I and sketched in Fig. 8.
Such an updating process is based on the degree of \emph{reliability}
of each $p$-th ($p=1,...,P$) trial solution, $\Psi_{i}^{\left(p\right)}$,
and of its previous best position, $\Psi\left\{ \underline{b}_{i-1}^{\left(p\right)}\right\} $.
For instance, let us consider the case illustrated in Fig. 8(\emph{d})
where the cost function value assigned to $\underline{\Omega}_{i}^{\left(p\right)}$,
$\widetilde{\Phi}_{i}^{\left(p\right)}$, and $\underline{b}_{i-1}^{\left(p\right)}$,
$\widetilde{\Phi}\left\{ \underline{b}_{i-1}^{\left(p\right)}\right\} $,
is affected by some uncertainty, both being predicted. Although $\widetilde{\Phi}_{i}^{\left(p\right)}>\widetilde{\Phi}\left\{ \underline{b}_{i-1}^{\left(p\right)}\right\} $,
it is profitable to update the $p$-th ($p=1,...,P$) personal best
(i.e., $\underline{b}_{i}^{\left(p\right)}\leftarrow\underline{\Omega}_{i}^{\left(p\right)}$)
since $\underline{\Omega}_{i}^{\left(p\right)}$ has a higher \emph{probability}
to have a smaller cost function value than $\underline{b}_{i-1}^{\left(p\right)}$
{[}i.e., $\mathcal{F}^{-}\left\{ \underline{\Omega}_{i}^{\left(p\right)}\right\} <\mathcal{F}^{-}\left\{ \underline{b}_{i-1}^{\left(p\right)}\right\} $
- Tab I and Fig. 8(\emph{d}){]}. Otherwise, whether the previous personal
best has been simulated {[}i.e., it is $100\%$ reliable - Fig. 8(\emph{b}){]},
it can be updated with the current particle position only if this
latter has no chance to have a worse/higher cost function value (i.e.,
$\mathcal{F}^{+}\left\{ \underline{\Omega}_{i}^{\left(p\right)}\right\} <\Phi\left\{ \underline{b}_{i-1}^{\left(p\right)}\right\} $
, $\mathcal{F}^{+}\left\{ \underline{\Omega}_{i}^{\left(p\right)}\right\} $
being the {}``upper confidence level'' (\emph{UCB}) associated to
$\underline{\Omega}_{i}^{\left(p\right)}$ given by $\mathcal{F}^{+}\left\{ \underline{\Omega}_{i}^{\left(p\right)}\right\} \triangleq\widetilde{\Phi}_{i}^{\left(p\right)}+\zeta\Psi_{i}^{\left(p\right)}$
{[}Tab. I - Fig. 8(\emph{b}){]};
\item Update the global best position, $\underline{g}_{i}$ ($\underline{g}_{i}\equiv\underline{\Omega}_{i}^{best}$)
according to the \emph{confidence-based} rules in Tab. II and illustrated
in Fig. 9. 
\item If $i=I$, then stop the optimization and output $\underline{g}_{I}$
as the final design (i.e., $\underline{\Omega}^{\left(end\right)}=\underline{g}_{I}$),
else go to Step 2(\emph{f});
\item Use the standard \emph{PSO} governing equations \cite{Rocca 2009}
to generate a new set of particles positions and velocities, update
the iteration index, $i\leftarrow\left(i+1\right)$, and go to Step
2(a).
\end{enumerate}
\end{enumerate}
As it can be inferred, the \emph{PSO-OK/C} \emph{SbD} method implements
a surrogate-assisted evolutionary optimization by exploiting an {}``individual-based''
\emph{model management} strategy \cite{Jin 2011}, but unlike state-of-the-art
techniques, it gives the user a full control of the time saving $\Delta t_{sav}$
by letting him specify the total amount $S$ of feasible \emph{FW}
simulations {[}i.e., $S=P\times I\times\left(1-\frac{\Delta t_{sav}}{100}\right)${]}
in order to comply with specific computational/time constraints for
the synthesis problem at hand.

\section{\emph{SbD} as Applied to the Synthesis of Complex \emph{EM} Systems}

This Section is aimed at assessing the effectiveness, the potentialities,
and the current limitations of the \emph{SbD} as applied to the design
of complex \emph{EM} systems. Towards this end, two representative
benchmarks, concerned with the synthesis of (\emph{i}) time modulated
arrays (\emph{TMA}s) (Sect. \ref{sub:Synthesis-of-Realistic-TMA})
and of (\emph{ii}) microstrip patch arrays for 5\emph{G} applications
(Sect. \ref{sub:Design-of-a-5G-array}), will be discussed.

\subsection{\emph{Benchmark 1} - Synthesis of \emph{TMA}s \label{sub:Synthesis-of-Realistic-TMA}}

The first benchmark problem deals with the synthesis of \emph{TMA}s
comprising real radiators and non-linear switching beam-forming networks
(\emph{BFN}s). To model the mutual coupling effects among the $N$
antennas as well as the complex nonlinear/dynamic behavior of the
\emph{BFN}, the Harmonic-Balance (\emph{HB}) technique \cite{Masotti 2019}\cite{Rizzoli 2011}
has been used. The goal is to determine the optimal setup of the switch-on
instants, $\underline{t}^{on}=\left\{ t_{n}^{on};\, n=1,...,N\right\} $,
$t_{n}^{on}$ being the $T$-normalized rise time of the $n$-th ($n=1,...,N$)
element subject to the physical-admissibility constraint $0\leq t_{n}^{on}<1$
(\emph{RCD} Block), that minimizes the fluctuations of the instantaneous
directivity $D$ within the modulation period $T$ \cite{Masotti 2019}.
Following the guidelines in Sect. \ref{sub:Solution-Model-and-DoF}
(\emph{PF} Block), the cardinality of the solution space (i.e., the
number of \emph{DoF}s) has been reduced to $K=\frac{N}{2}$ by considering
symmetric excitation sequences {[}i.e., $t_{n}^{on}=t_{\left(N-n+1\right)}^{on}$;
$n=\left(\frac{N}{2}+1\right),....,N${]} so that $\underline{\Omega}=\left\{ t_{n}^{on};\, n=1,...,\frac{N}{2}\right\} $,
while the cost function has been defined as\begin{equation}
\Phi\left\{ \underline{\Omega}\right\} =\frac{1}{\overline{D}\left\{ \underline{\Omega}\right\} T}\int_{0}^{T}\left|\overline{D}\left\{ \underline{\Omega}\right\} -D\left\{ \underline{\Omega},\, t_{p}\right\} \right|dt_{p}\label{eq:cost-function-TMA}\end{equation}
 $\overline{D}$ being the average value of the instantaneous directivity
over $T$. 

\noindent As a representative numerical test, a \emph{TMA} with $N=16$
monopoles resonating at $f_{0}=2.45$ {[}GHz{]} printed over a \emph{RF60-A}
Taconic substrate with relative permittivity $\varepsilon_{r}=6.15$,
loss tangent $\tan\delta=0.0028$, and thickness $h=0.635$ {[}mm{]}
\cite{Masotti 2019} has been considered. Moreover, the duration of
the modulation period has been set to $T=10$ {[}$\mu sec${]}, while
the durations of the driving pulses, $\underline{\tau}=\left\{ \tau_{n};\, n=1,\,...,\, N\right\} $,
have been chosen to afford a Dolph-Chebyshev pattern with a side-lobe
level of $SLL=-30$ {[}dB{]} at $f_{0}$.

\noindent In order to define the most suitable implementation of the
\emph{CFC} and \emph{SSE} blocks, a preliminary study has been carried
out to (\emph{i}) estimate the nature of the cost function (\ref{eq:cost-function-TMA})
as well as to (\emph{ii}) assess the accuracy of different \emph{SM}s.
More specifically, the plots of the 1\emph{-D} cuts of $\Phi$ (\ref{eq:functional-cut})
have been evaluated and an example is reported in Fig. 10(\emph{a}).
In this latter, the trial solution $\underline{\Omega}=\underline{\Omega}^{\left(1\right)}$
($t=0$) corresponds to a randomly-chosen position within the search
space $\mathcal{\Re}\left\{ \underline{\Omega}\right\} $, while $\underline{\Omega}=\underline{\Omega}^{\left(2\right)}$
($t=1$) is the solution found by a state-of-the-art \emph{StD} approach
based on a \emph{PSO} run of $I=200$ iterations with $P=10$ particles
\cite{Masotti 2019} {[}$\Phi_{StD}=3.2\times10^{-2}$ ($\Phi_{StD}\triangleq\Phi\left\{ \underline{\Omega}_{StD}^{\left(end\right)}\right\} $)
- Fig. 10(\emph{a}){]}. As it can be observed, the cost function (\ref{eq:cost-function-TMA})
has a highly-oscillating multi-modal behavior with the occurrence
of many local minima and the presence of steep/non-symmetric valleys
{[}Fig. 10(\emph{a}){]}. Those features imply that the \emph{SSE}
block cannot be a deterministic optimizer, but a stochastic hill-climbing
techniques is mandatory. 

\noindent Next, the accuracy of different \emph{SM}s (namely, the
\emph{RBFN}, the \emph{SVR}, and the \emph{OK} models) has been preliminary
analyzed by checking the dependence of the prediction accuracy on
the number of training samples, $S$. Towards this aim, a set of $U=\left(P\times I\right)=2000$
\emph{HB-FW} simulations has been performed by \emph{LHS}-sampling
the solution space, then the \emph{SM}s have been trained with a subset
of $S<U$ samples and their accuracy has been evaluated on the remaining
$M=\left(U-S\right)$ samples. Figure 10(\emph{b}) reports the values
of the \emph{SM} error\begin{equation}
\Xi\left(S\right)=\frac{\sum_{m=1}^{M=\left(U-S\right)}\left|\widetilde{\Phi}\left\{ \left.\underline{\Omega}^{\left(m\right)}\right|S\right\} -\Phi\left\{ \underline{\Omega}^{\left(m\right)}\right\} \right|^{2}}{\sum_{m=1}^{M=\left(U-S\right)}\left|\Phi\left\{ \underline{\Omega}^{\left(m\right)}\right\} \right|^{2}}\label{eq:}\end{equation}
versus $S$ along with the time saving $\Delta t_{sav}$ of the \emph{SbD}
{[}Fig. 10(\emph{b}){]}. As expected, $\Delta t_{sav}$ (linearly)
decreases with $S$ independently on the adopted \emph{SM}, while
the \emph{OK}-based surrogate always outperforms the other \emph{SM}s
in terms of prediction performance, the \emph{SVR} generally providing
the worse values of $\Xi$ {[}Fig. 10(\emph{b}){]}. It is also worth
pointing out that, no matter what the training set size within the
range $0.1\le\frac{S}{U}\le0.75$, the value of $\Xi$ of each method
is almost constant and smaller than $10^{-1}$.

\noindent Moving to the solution of the \emph{TMA} design problem
with standard \emph{SbD} approaches using \emph{LHS} training sets
of different sizes, let us analyze the case of very high time saving
(i.e., $\Delta t_{sav}\geq90\%$ $\rightarrow$ $S\leq200$). By computing
the {}``design quality index'', $\Delta\Phi\triangleq\frac{\left(\Phi_{SbD}-\Phi_{StD}\right)}{\Phi_{StD}}$,
as the normalized difference between the cost function of the \emph{StD}
solution, $\Phi_{StD}$, and that from the \emph{SbD}, $\Phi_{SbD}\triangleq\Phi\left\{ \underline{\Omega}_{SbD}^{\left(end\right)}\right\} $,
it turns out that both the \emph{DE-OK} and the \emph{PSO-OK} always
yield the best results {[}Fig. 10(\emph{c}){]}. The performance of
the \emph{advanced} \emph{SbD} approaches presented in Sect. \ref{sec:Advanced-SbD-Strategies}
have been assessed, as well. They allows one remarkable improvements
with respect to all bare \emph{SbD} implementations. Indeed, Figure
11 indicates that the {}``smart'' sampling performed by the \emph{SbD-OSF}
technique (here initialized with $S_{0}=40$ samples) results in a
significant reduction of $\Delta\Phi$ when applying, later on, the
\emph{PSO-OK}. Even more significant is the positive effect of the
\emph{PSO-OK/C} (still initialized with $S_{0}=40$ samples). For
instance, the \emph{PSO-OK/C} obtains a solution very close to that
of the \emph{StD} approach when $S=100$ (i.e., $\left.\Delta\Phi\right|_{PSO-OK/C}^{S=100}=6.2\%$
- Fig. 11), but with an impressive time saving of $\Delta t_{sav}=95\%$.
In order to better appreciate such an outcome, let us consider that
$\Delta t_{FW}=13$ {[}sec{]} on a desktop PC with 3.6 GHz CPU and
32 GB of RAM memory, thus the time required by the \emph{StD} approach
and the \emph{PSO-OK/C} to complete the optimization ($i=I$) was
$\Delta t_{StD}=7.2$ {[}hours{]} and $\left.\Delta t\right|_{PSO-OK/C}^{S=100}=21.2$
{[}min{]}, respectively.

\noindent The effectiveness of the \emph{PSO-OK/C} in exploring the
solution space are confirmed by the evolution of the optimal value
of the cost function during the iterative optimization process ($i=1,...,I$)
{[}$\Phi_{i}^{best}$ - Fig. 12(\emph{a}){]} as well as by the behavior
of the optimized instantaneous directivities {[}Fig. 12(\emph{b}){]},
which appears very similar to that of the \emph{StD} approach in both
plots. It is also interesting to note the progressive refinement of
the \emph{OK} surrogate during the \emph{PSO-OK/C} optimization as
it can be inferred from the plot of the functional cut passing through
the global best solutions at $i=0$ (i.e., $\underline{g}_{i=0}$
- $t=0$) and $i=I$ ($\underline{g}_{i=I}$ - $t=1$) (Fig. 13).
As a matter of fact, there is a non-negligible improvement of the
\emph{SM} accuracy in mapping the actual behavior of the cost function
from the initial ($i=0$) to the final iteration ($i=I$) thanks to
the adaptive selection, during the optimization, of new training samples
(Fig. 13).

\noindent As for the improvement of the \emph{PSO-OK/C} over the \emph{PSO-OK},
Figure 14 compares the predicted values of the cost function, $\widetilde{\Phi}\left\{ \underline{g}_{j}\right\} $,
with the corresponding actual ones, $\Phi\left\{ \underline{g}_{j}\right\} $,
at some intermediate {}``control-points'' ($j=1,...,J$; $J$ being
the maximum number of selected iterations). Unlike the \emph{PSO-OK}
(although the same number of $S=100$ \emph{FW} simulations has been
used), there is a perfect matching between the \emph{PSO-OK/C} predictions
and the actual cost function values thanks to the adaptive addition
of training samples during the minimization. Quantitatively, it turns
out that $\Phi_{SbD\,(PSO-OK)}=9.4\times10^{-2}$ vs. $\Phi_{SbD\,(PSO-OK/C)}=3.4\times10^{-2}$
(Fig. 14).

\noindent Finally, the \emph{SbD} has been compared with competitive
state-of-the-art approaches. First, the surrogate-assisted \emph{DE}
(\emph{SADE}) algorithm \cite{Liu 2014} has been taken into account
and Figure 11 proves that the \emph{PSO-OK/C} positively compares
with such an optimization approach always providing good trade-offs
between design-quality and time saving. Moreover, a comparison between
the \emph{PSO-OK/C} performance and those from a space mapping technique
\cite{Bandler 2004} has been carried out, as well. Towards this end,
the additive-input/multiplicative-output (\emph{AIMO}) space mapping
approach \cite{Koziel 2009} has been implemented by considering an
analytic model for the \emph{TMA} composed by isotropic radiators
and ideal switches ($\Delta t_{coarse}=0.02$ {[}sec{]}). It turns
out that, despite a lightly better time saving with just $S=6$ \emph{HB-FW}
simulations and $S_{coarse}=2200$ coarse-model evaluations (i.e.,
$\left.\Delta t_{sav}\right|_{AIMO}=97.9\%$), the final solution
is slightly less accurate than the \emph{PSO-OK/C} one (i.e., $\Phi_{AIMO}=5.9\times10^{-2}$).

\subsection{\emph{Benchmark 2} - Design of a Microstrip Array for 5\emph{G} Applications
\label{sub:Design-of-a-5G-array}}

\noindent This second test case is devoted to further assess the advantages
of the \emph{SbD} when dealing with \emph{high-complexity} \emph{EM}
design problems. Let us consider the synthesis of a planar microstrip
array for 5\emph{G} applications working at $f_{0}=3.5$ {[}GHz{]}
(Fig. 15). The antenna has been supposed to lie on the $\left(y,\, z\right)$-plane
and composed by $N=\left(4\times6\right)$ cavity-backed slot-fed
square patches of side $L_{p}=2.73\times10^{-1}$ $\lambda$. To enable
$\pm45$ {[}deg{]} dual-polarization operation ($J=2$ being the number
of operating modes), each element has been rotated by $45$ {[}deg{]}
and fed by two rectangular slots of dimensions $\left(W_{h},\, L_{h}\right)=\left(2.65\times10^{-2},\,1.31\times10^{-1}\right)$
$\lambda$ {[}Fig. 15(\emph{b}){]}. The dimensions of the feeding
lines have been set to $\left(W_{f},\, L_{f}\right)=\left(1.69\times10^{-2},\,1.17\times10^{-1}\right)$
$\lambda$, with an additional stub of length $L_{s}=5.25\times10^{-2}$
$\lambda$ to reach a proper impedance matching. The following material
and dielectric properties have been assumed: Arlon DiClad527 ($\varepsilon_{r}=2.5$,
$\tan\delta=0.0018$) with a total thickness of $H=\left(h_{1}+h_{2}+h_{c}\right)=1.24\times10^{-1}$
$\lambda$ {[}Fig. 15(\emph{a}){]}, $h_{1}=1.77\times10^{-2}$ $\lambda$,
$h_{2}=5.93\times10^{-3}$ $\lambda$, and $h_{c}=1.0\times10^{-1}$
$\lambda$ being the height of the patch substrate, of the feeding
substrate, and of the cavity, respectively. Starting from this reference
setup, the set of \emph{DoF}s $\underline{\Omega}=\left\{ d_{y},\, d_{z},\, l_{1},\, l_{2}\right\} $
($K=4$), $d_{y}$ and $d_{z}$ being the inter-element spacing along
$y$ and $z$, respectively, while $l_{1}$ and $l_{2}$ locate the
position of the feeding slot and microstrip line associated to polarization/port
$j=2$ {[}Fig. 15(\emph{b}){]}, have been optimized to maximize the
realized gain along the main beam direction $\left(\theta_{o},\,\varphi_{o}\right)$,
$G_{j,o}\left(\theta_{o},\,\varphi_{o}\right)$, for both polarizations
and also towards $O=5$ steering angles ($o=1,...,O$) within the
visible cone of the array %
\footnote{\noindent More in detail, the following steering directions have been
considered: $\left(\theta_{1},\,\varphi_{1}\right)=\left(90,\,0\right)$
{[}deg{]}, $\left(\theta_{2},\,\varphi_{2}\right)=\left(-60,\,75\right)$
{[}deg{]}, $\left(\theta_{3},\,\varphi_{3}\right)=\left(60,\,75\right)$
{[}deg{]}, $\left(\theta_{4},\,\varphi_{4}\right)=\left(-60,\,105\right)$
{[}deg{]}, and $\left(\theta_{5},\,\varphi_{5}\right)=\left(60,\,105\right)$
{[}deg{]}.%
}. Mathematically, such a synthesis problem has been coded (\emph{PF}
Block) into the following cost function\begin{equation}
\Phi\left\{ \underline{\Omega}\right\} =\left(\frac{1}{2\times O}\sum_{j=1}^{J}\sum_{o=1}^{O}G_{j,o}\left(\left.\theta_{o},\,\varphi_{o}\right|\underline{\Omega}_{5G}\right)\right).^{-2}\label{eq:cost-function-5G}\end{equation}
As for the computation of the actual value of the cost function $\Phi$,
the \emph{FW} problem at hand has been solved with the Ansys \emph{HFSS}
\emph{EM} simulator \cite{HFSS} by considering exactly the \emph{finite}
structure of the antenna (i.e., no periodic-infinite hypotheses) to
take into account both the mutual coupling effects among the array
elements as well as the fringing effects, the average simulation time
being equal to $\Delta t_{FW}=5.64\times10^{3}$ {[}sec{]}.

\noindent Figure 16 shows the evolution of the optimal value of the
cost function for a \emph{StD} optimization based on the \emph{PSO}
executed with $P=4$ agents for $I=50$ iterations ($P\times I=200$
$\rightarrow$ $\Delta t_{StD}=1.13\times10^{6}$ {[}sec{]} $\sim$
$13$ {[}days{]}), $\Phi\left\{ \underline{g}_{i}\right\} $ ($i=1,...,I$),
as well as the predicted curve outputted by the \emph{PSO-OK/C}, $\Phi_{i}$,
along with the intermediate {}``control points''. The \emph{SbD}
converges to a solution whose cost function value is only $\Delta\Phi=2.64\%$
greater than that of the \emph{StD}, but with a remarkable advantage
in terms of computational efficiency. As a matter of fact, the \emph{PSO-OK/C}
performed only $S=40$ \emph{FW} simulations (comprising $S_{0}=20$
initial training samples before the optimization) enabling a time
saving of $\Delta t_{sav}=80\%$. For completeness, the plot of the
arising realized gain patterns are shown in Fig. 17 when steering
the main beam towards $\left(\theta_{o},\,\varphi_{o}\right)=\left(90,\,0\right)$
{[}deg{]} {[}Fig. 17(\emph{a}){]} and $\left(\theta_{o},\,\varphi_{o}\right)=\left(105,\,60\right)$
{[}deg{]} {[}Fig. 17(\emph{b}){]}. For completeness, the average realized
gains along the steering direction are very close {[}i.e., $\left.\overline{D}_{StD}\right\rfloor _{j=1}=15.43$
{[}dB{]} vs. $\left.\overline{D}_{SbD}\right\rfloor _{j=1}=15.22$
{[}dB{]} and $\left.\overline{D}_{StD}\right\rfloor _{j=2}=15.53$
{[}dB{]} vs. $\left.\overline{D}_{SbD}\right\rfloor _{j=2}=15.36$
{[}dB{]}, being $\overline{D}_{j}\triangleq\frac{1}{2\times O}\sum_{o=1}^{O}G_{j,o}\left(\left.\theta_{o},\,\varphi_{o}\right|\underline{\Omega}\right)${]},
further confirming the similarity of the solutions at the convergence.

\section{\noindent Conclusions, Final Remarks, and Future Trends}

The \emph{SbD} is an innovative paradigm for the computationally-efficient
solution of complex \emph{EM} synthesis problems mainly devoted to
properly deal with the {}``high complexity'' curse. Towards this
purpose, the synthesis problems at hand are addressed through a suitable
problem-driven selection and interconnection of \emph{functional blocks},
each one implementing a rather {}``simple'' and well-identified
task, but all integrated to jointly fit, in an easier fashion, the
required system functionality/performance. From a theoretical point
of view, after summarizing the key-features and the building blocks
of the \emph{SbD}-based synthesis framework, two innovative \emph{SbD}
implementations have been presented and applied to two challenging
problems concerned with the design of realistic \emph{TMA}s (Sect.
\ref{sub:Synthesis-of-Realistic-TMA}) and the synthesis of planar
dual-polarization microstrip arrays for 5\emph{G} applications (Sect.
\ref{sub:Design-of-a-5G-array}).

\noindent The main outcomes from these analyses are as follows: 

\begin{itemize}
\item the \emph{SbD} enables the design of complex \emph{EM} devices/systems
in a suitable time-frame thanks to the possibility to select the best
(i.e., problem-oriented) trade-off between time saving and prediction
accuracy;
\item the accurate selection of the solution descriptors (i.e., the \emph{DoF}s)
allows one to reduce the number of required training samples and therefore
the computational burden of the training phase;
\item the role of the \emph{SM}s in the \emph{SbD} is not only that of a
reliable (i.e., accurate prediction of the cost function values) and
computationally-cheap alternative to the \emph{FW} solver, but they
mainly devoted to map the landscape of the cost function to reliably
drive the search for the global optimum of the \emph{SSE} algorithm.
\end{itemize}
Future activities, out of the scope of this work, will be aimed at
customizing and applying the \emph{SbD} to further challenging and
high-complexity designs including unconventional phased arrays with
real elements \cite{Rocca 2016}, and innovative meta-materials for
smart \emph{EM} environments applications \cite{Basar 2019}. Moreover,
the development of \emph{SbD} strategies, which exploit deep learning
\cite{Massa 2019} and/or multi-objective optimization algorithms
\cite{Kadlec 2013}, is under investigation. Finally, innovative optimization-driven
methodologies based on the Compressive Sensing (\emph{CS}) paradigm
will be studied to build optimal training sets overcoming the Nyquist's
theoretical limit so that it would possible to faithfully predict
the cost function values using far fewer samples than traditional
approaches \cite{Massa 2015}.

\section*{Acknowledgements}

\noindent This work has been partially supported by the Italian Ministry
of Education, University, and Research within the Program \char`\"{}Smart
cities and communities and Social Innovation\char`\"{} (CUP: E44G14000060008)
for the Project \char`\"{}WATERTECH - Smart Community per lo Sviluppo
e l'Applicazione di Tecnologie di Monitoraggio Innovative per le Reti
di Distribuzione Idrica negli usi idropotabili ed agricoli\char`\"{}
(Grant no. SCN\_00489) and by the Ministry of Education of China within
the Chang-Jiang Visiting Professor chair. A. Massa wishes to thank
E. Vico for her never-ending inspiration, support, guidance, and help.

\newpage
\section*{FIGURE CAPTIONS}

\begin{itemize}
\item \textbf{Figure 1.} \emph{SbD Paradigm} - Functional scheme of the
\emph{SbD}.
\item \textbf{Figure 2.} \emph{Problem Formulation -} Sketch of the (\emph{a})
pixel-based and (\emph{b}) spline-based representation of the \emph{TO}
contour $\gamma\left(x,\, y\right)$ \cite{Salucci 2019}.
\item \textbf{Figure 3.} \emph{Training Set Generation} - Pictorial representation
of the \emph{DoF}s, reduced features, and cost function spaces and
their interconnections for the generation of the training sets $\mathcal{D}_{S}=\left\{ \left(\underline{\Omega}^{\left(s\right)};\,\Phi^{\left(s\right)}\right);\, s=1,...,S\right\} $
and/or $\widehat{\mathcal{D}}_{S}=\left\{ \left(\underline{\chi}^{\left(s\right)};\,\Phi^{\left(s\right)}\right);\, s=1,...,S\right\} $.
\item \textbf{Figure 4.} \emph{Cost Function Computation} - Actual function
value and predictions made by the \emph{RBFN}, \emph{SVR}, and \emph{OK}
surrogate models for the (\emph{a}) Levy's, (\emph{b}) the Schwefel's,
and (\emph{c}) the Ackley's 1-\emph{D} ($K=1$) benchmark functions
when using $S=6$ training samples.
\item \textbf{Figure 5.} \emph{Cost Function Computation} - Actual, $\Phi\left\{ \underline{\Omega}\right\} $,
predicted, $\widetilde{\Phi}\left\{ \underline{\Omega}\right\} $,
and confidence bounds, $\widetilde{\Phi}\left\{ \underline{\Omega}\right\} \pm\Psi\left\{ \underline{\Omega}\right\} $,
of the \emph{OK} surrogate model.
\item \textbf{Figure 6.} \emph{Solution Space Exploration} - Values of the
cost function at the convergence, $\Phi_{StD}$, for different {}``bare''
\emph{SbD} algorithms versus the ratio between the training cardinality
and the number of variables, $S/K$: (\emph{a}) Levy's, (\emph{b})
Schwefel's, and (\emph{c}) Ackley's functions with $K=6$.
\item \textbf{Figure 7.} \emph{Advanced} \emph{SbD Strategies} (\emph{SbD-OSF};
Ackley's function) - (\emph{a}) Training samples generated by the
initial \emph{LHS} sampling ($S_{0}=5$) along with those iteratively
added by the \emph{SbD-OSF} ($S=50$); (\emph{b}) \emph{OK} predictions
with the initial (\emph{LHS}) or the \emph{SbD-OSF} training set.
\item \textbf{Figure 8.} \emph{Advanced} \emph{SbD Strategies} (\emph{SbD
PSO-OK/C}) - Pictorial sketch of the updating rules at the $i$-th
($i=1,...,I$) iteration for the personal best position of each $p$-th
($p=1,...,P$) particle, $\underline{b}_{i}^{\left(p\right)}$ ($\underline{b}_{i}^{\left(p\right)}=\arg\left\{ \min_{j=1,...,i}\left[\Phi_{j}^{\left(p\right)}\right]\right\} $).
\item \textbf{Figure 9.} \emph{Advanced} \emph{SbD Strategies} (\emph{SbD
PSO-OK/C}) - Pictorial sketch of the \emph{confidence-based} updating
rules at the $i$-th ($i=1,...,I$) iteration for the global best
position, $\underline{g}_{i}$ ($\underline{g}_{i}\equiv\underline{\Omega}_{i}^{best}$).
\item \textbf{Figure 10.} \emph{SbD-Synthesis} (\emph{Benchmark 1}: $K=8$,
$P=10$, $I=200$) - (\emph{a}) 1\emph{D} cut of $\Phi$; (\emph{b})
prediction error, $\Xi$, and time saving, $\Delta t_{sav}$, versus
the size of the training set, $S$; (\emph{c}) performance indexes
(i.e., $\Delta\Phi$ and $\Delta t_{sav}$) versus $S$.
\item \textbf{Figure 11.} \emph{SbD-Synthesis} (\emph{Benchmark 1}: $K=8$,
$P=10$, $I=200$) - Design quality index, $\Delta\Phi$, and SbD
time saving, $\Delta t_{sav}$, versus $S$.
\item \textbf{Figure 12.} \emph{SbD-Synthesis} (\emph{Benchmark 1}: $K=8$,
$P=10$, $I=200$, $S_{0}=40$, $S=100$) - (\emph{a}) Evolution of
the best value of the cost function, $\Phi_{i}^{best}$, versus the
iteration index, $i$ ($i=1,...,I$) and (\emph{b}) plot of the instantaneous
directivity over one modulation period $T$.
\item \textbf{Figure 13.} \emph{SbD-Synthesis} (\emph{Benchmark 1}: $K=8$,
$P=10$, $I=200$, $S_{0}=40$, $S=100$) - One-dimensional cut of
$\Phi\left\{ \underline{\Omega}\right\} $ and \emph{OK} predictions
when applying the \emph{PSO-OK/C.}
\item \textbf{Figure 14.} \emph{SbD-Synthesis} (\emph{Benchmark 1}: $K=8$,
$P=10$, $I=200$, $S_{0}=40$, $S=100$) - Evolution of $\widetilde{\Phi}\left\{ \underline{g}_{j}\right\} $
versus the iteration index, $i$ ($i=1,...,I$) and cost function
values at the $J$ control-iterations, \{$\Phi\left\{ \underline{g}_{j}\right\} $;
$j=1,...,J$\}.
\item \textbf{Figure 15.} \emph{SbD-Synthesis} (\emph{Benchmark 2}: $K=4$)
- Geometry of (\emph{a}) the complete antenna array system modeled
with \emph{HFSS} and (\emph{b}) details of the elementary radiator,
namely a cavity-backed slot-fed square patch.
\item \textbf{Figure 16} \emph{SbD-Synthesis} (\emph{Benchmark 2}: $K=4$,
$P=4$, $I=50$, $S_{0}=20$, $S=40$) - Evolution of the best value
of the cost function, $\Phi_{i}^{best}$, versus the iteration index,
$i$ ($i=1,...,I$).
\item \textbf{Figure 17.} \emph{SbD-Synthesis} (\emph{Benchmark 2}: $K=4$,
$P=4$, $I=50$, $S_{0}=20$, $S=40$) - Optimized realized gain patterns
for the $j$-th ($j=1,\,2$) polarization when steering the main beam
towards (\emph{a}) $\left(\theta_{0},\,\varphi_{0}\right)=\left(90,\,0\right)$
{[}deg{]} and (\emph{b}) $\left(\theta_{0},\,\varphi_{0}\right)=\left(105,\,60\right)$
{[}deg{]}.
\end{itemize}

\section*{TABLE CAPTIONS}

\begin{itemize}
\item \textbf{Table I.} \emph{Advanced} \emph{SbD Strategies} (\emph{SbD
PSO-OK/C}) - Personal best update rules.
\item \textbf{Table II.} \emph{Advanced} \emph{SbD Strategies} (\emph{SbD
PSO-OK/C}) - Global best update rules.\newpage

\end{itemize}
\begin{center}~\vfill\end{center}

\begin{center}\begin{tabular}{c}
\includegraphics[%
  width=0.90\columnwidth]{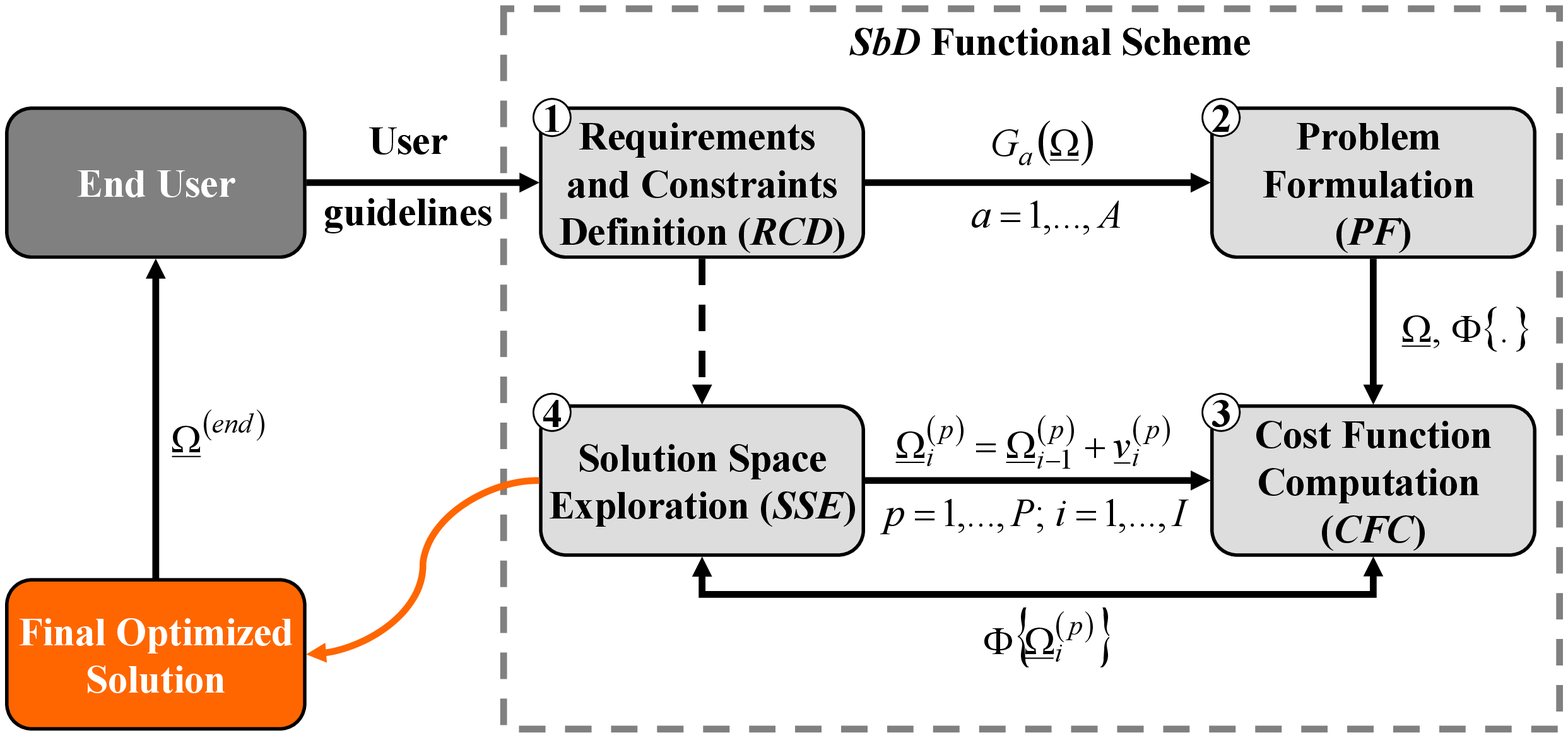}\tabularnewline
\end{tabular}\end{center}

\begin{center}~\vfill\end{center}

\begin{center}\textbf{Fig. 1 - A. Massa et} \textbf{\emph{al.}}\textbf{,}
\textbf{\emph{{}``}}On the Design of Complex ...''\end{center}
\newpage

\begin{center}~\vfill\end{center}

\begin{center}\begin{tabular}{c}
\includegraphics[%
  width=0.60\columnwidth]{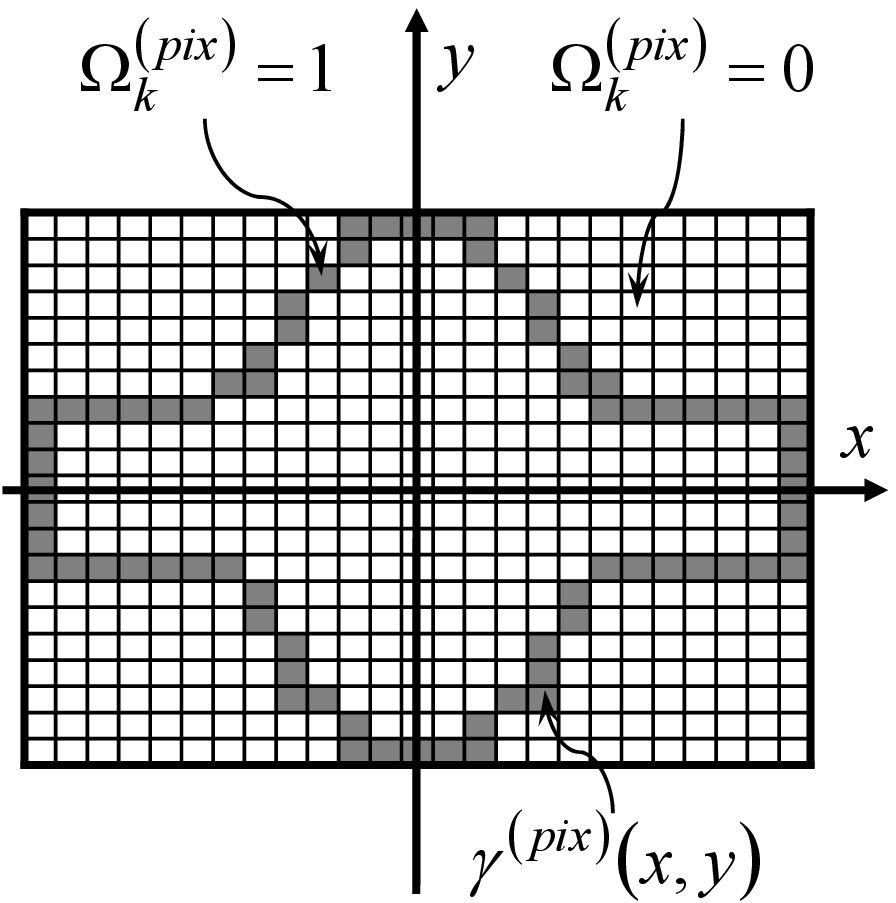}\tabularnewline
(\emph{a})\tabularnewline
\tabularnewline
\includegraphics[%
  width=0.60\columnwidth]{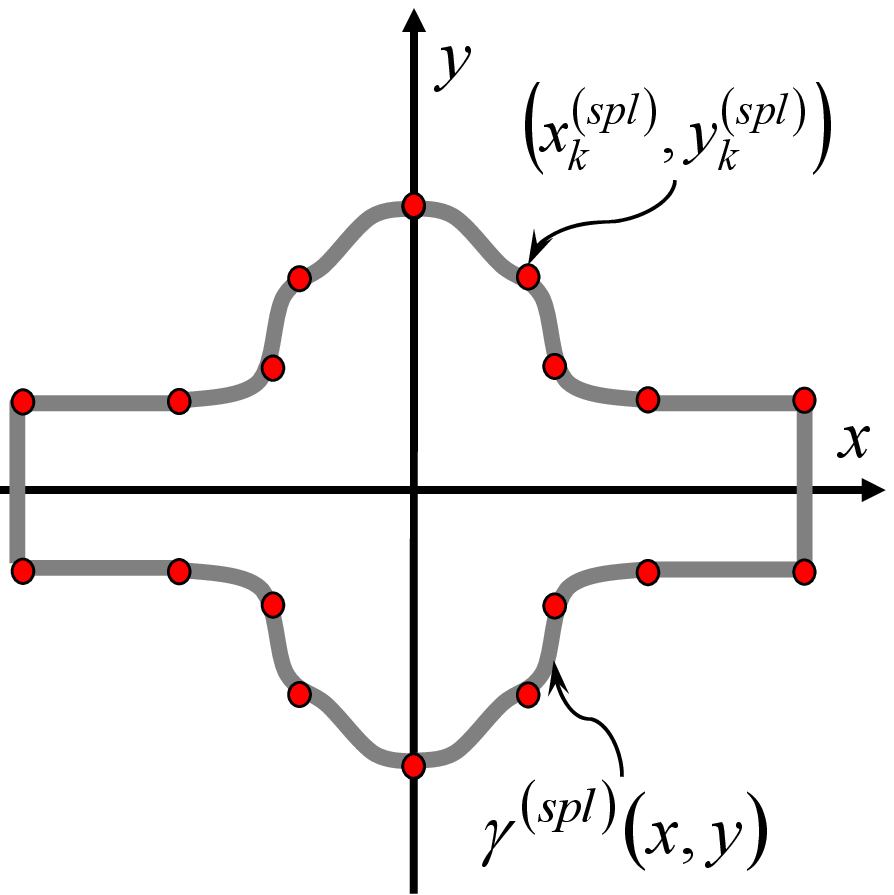}\tabularnewline
(\emph{b})\tabularnewline
\end{tabular}\end{center}

\begin{center}~\vfill\end{center}

\begin{center}\textbf{Fig. 2 - A. Massa et} \textbf{\emph{al.}}\textbf{,}
\textbf{\emph{{}``}}On the Design of Complex ...''\end{center}
\newpage

\begin{center}~\vfill\end{center}

\begin{center}\begin{tabular}{c}
\includegraphics[%
  width=1.0\columnwidth]{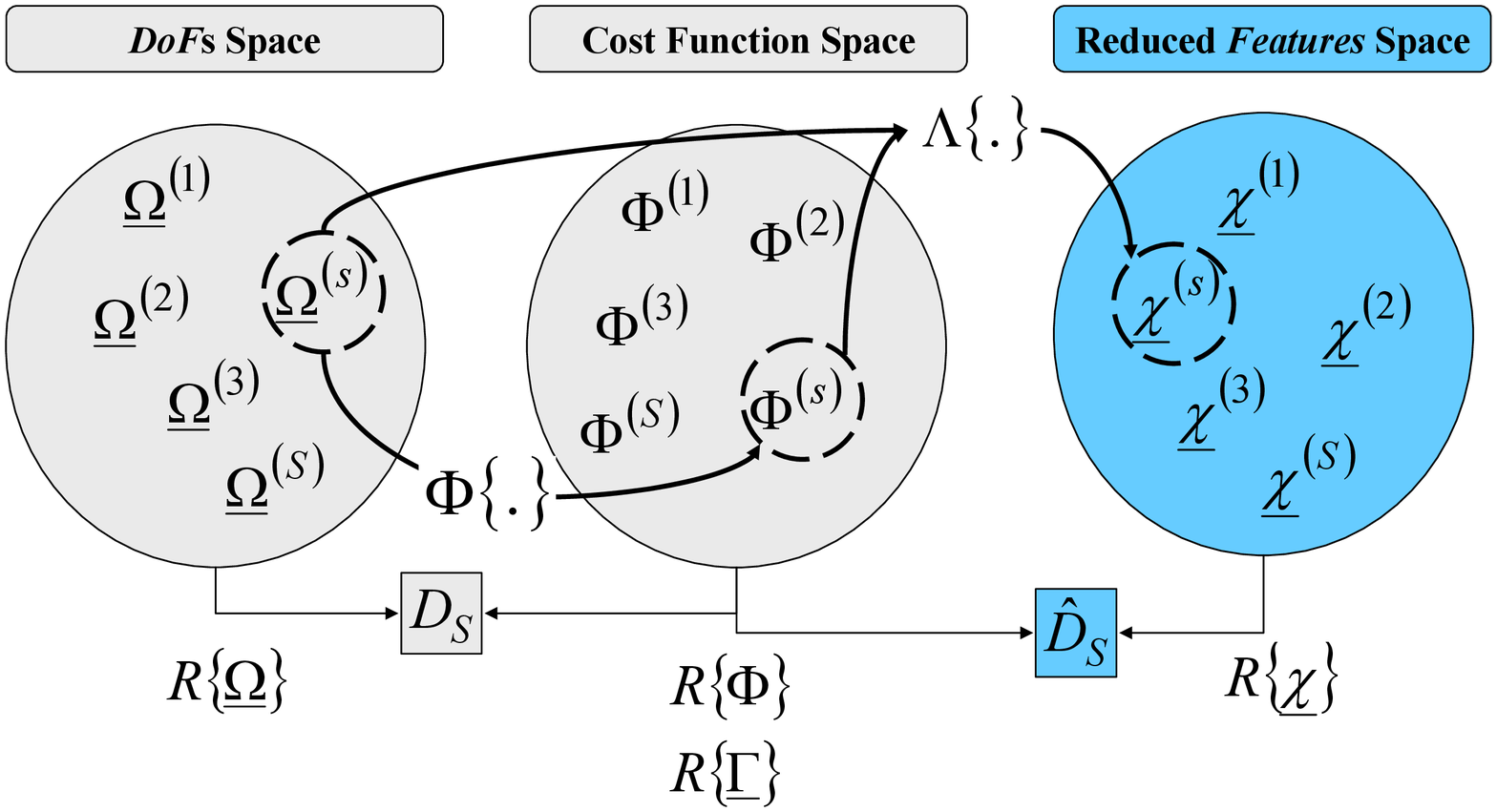}\tabularnewline
\end{tabular}\end{center}

\begin{center}~\vfill\end{center}

\begin{center}\textbf{Fig. 3 - A. Massa et} \textbf{\emph{al.}}\textbf{,}
\textbf{\emph{{}``}}On the Design of Complex ...''\end{center}
\newpage

\begin{center}~\vfill\end{center}

\begin{center}\begin{tabular}{c}
\includegraphics[%
  width=0.45\paperwidth]{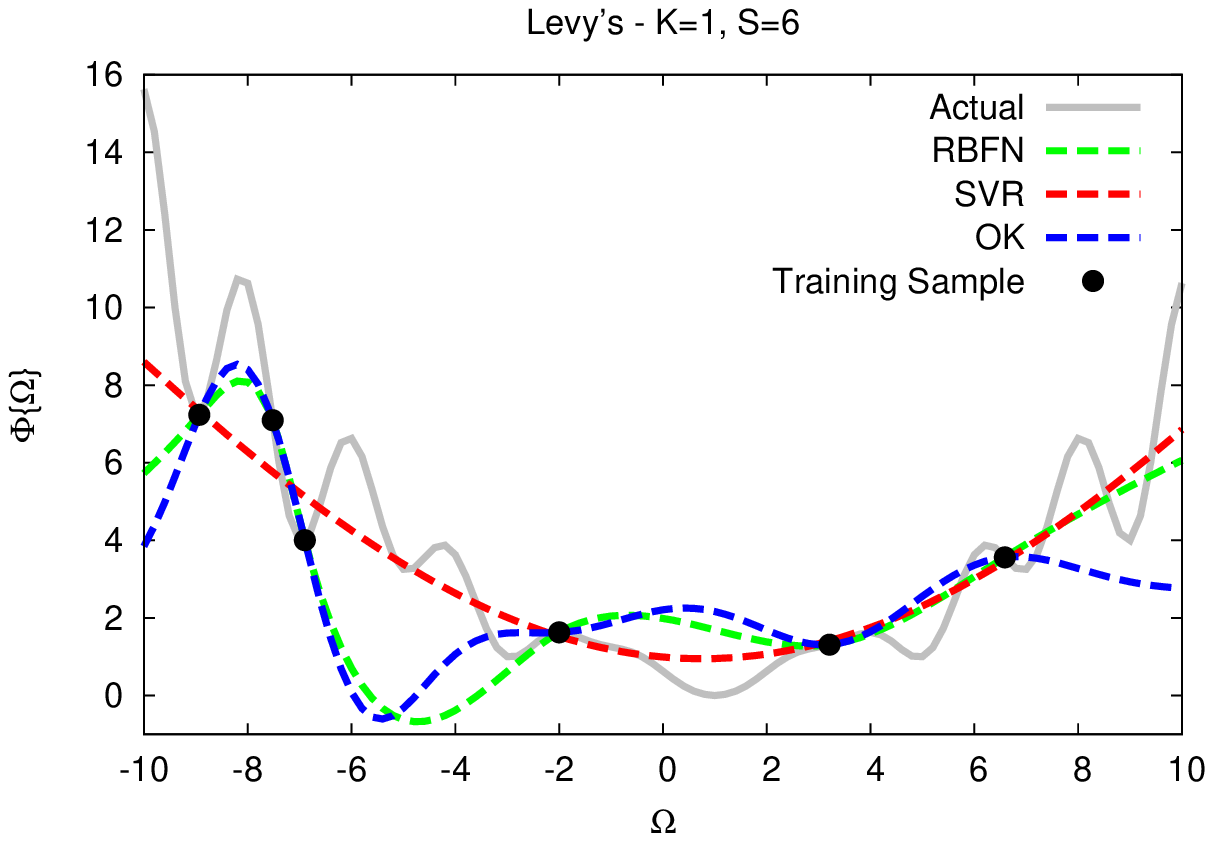}\tabularnewline
(\emph{a})\tabularnewline
\includegraphics[%
  width=0.45\paperwidth]{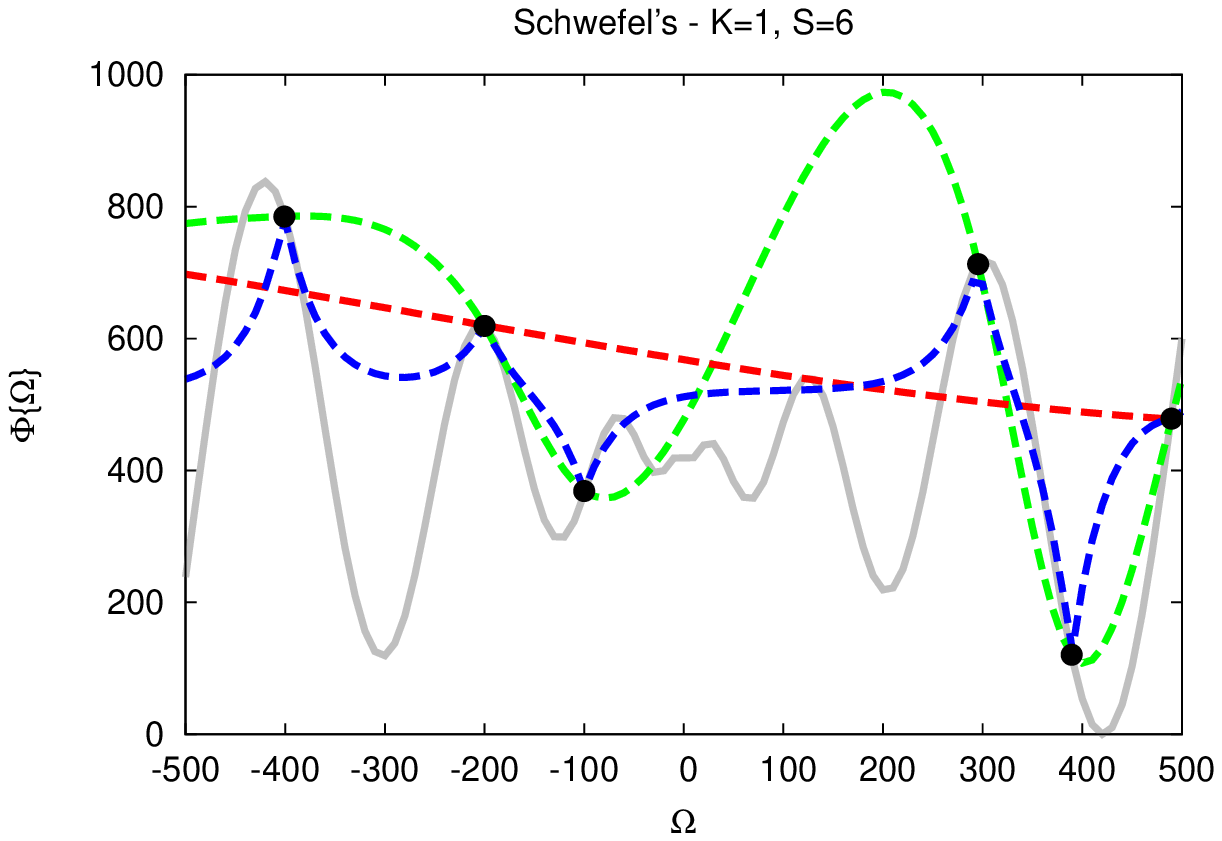}\tabularnewline
(\emph{b})\tabularnewline
\includegraphics[%
  width=0.45\paperwidth]{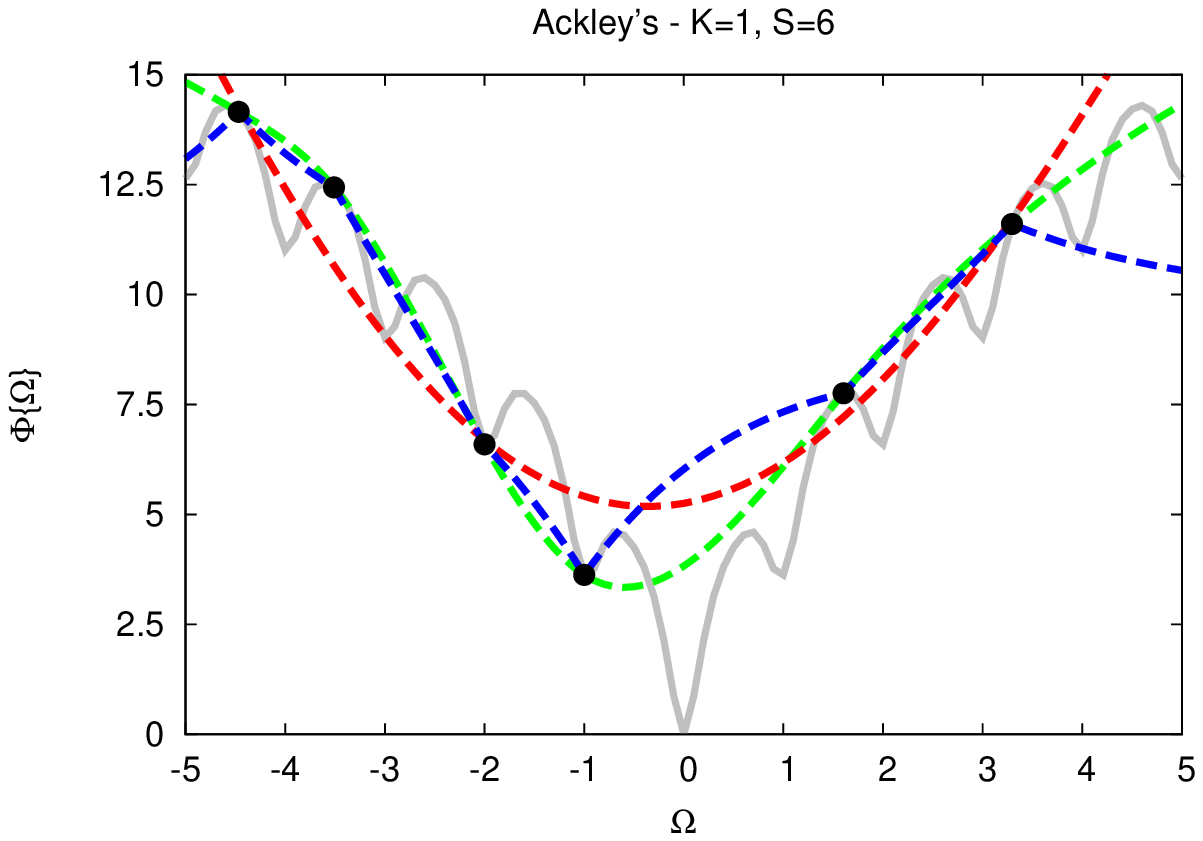}\tabularnewline
(\emph{c})\tabularnewline
\end{tabular}\end{center}

\begin{center}\textbf{Fig. 4 - A. Massa et} \textbf{\emph{al.}}\textbf{,}
\textbf{\emph{{}``}}On the Design of Complex ...''\end{center}
\newpage

\begin{center}~\vfill\end{center}

\begin{center}\begin{tabular}{c}
\includegraphics[%
  width=1.0\columnwidth]{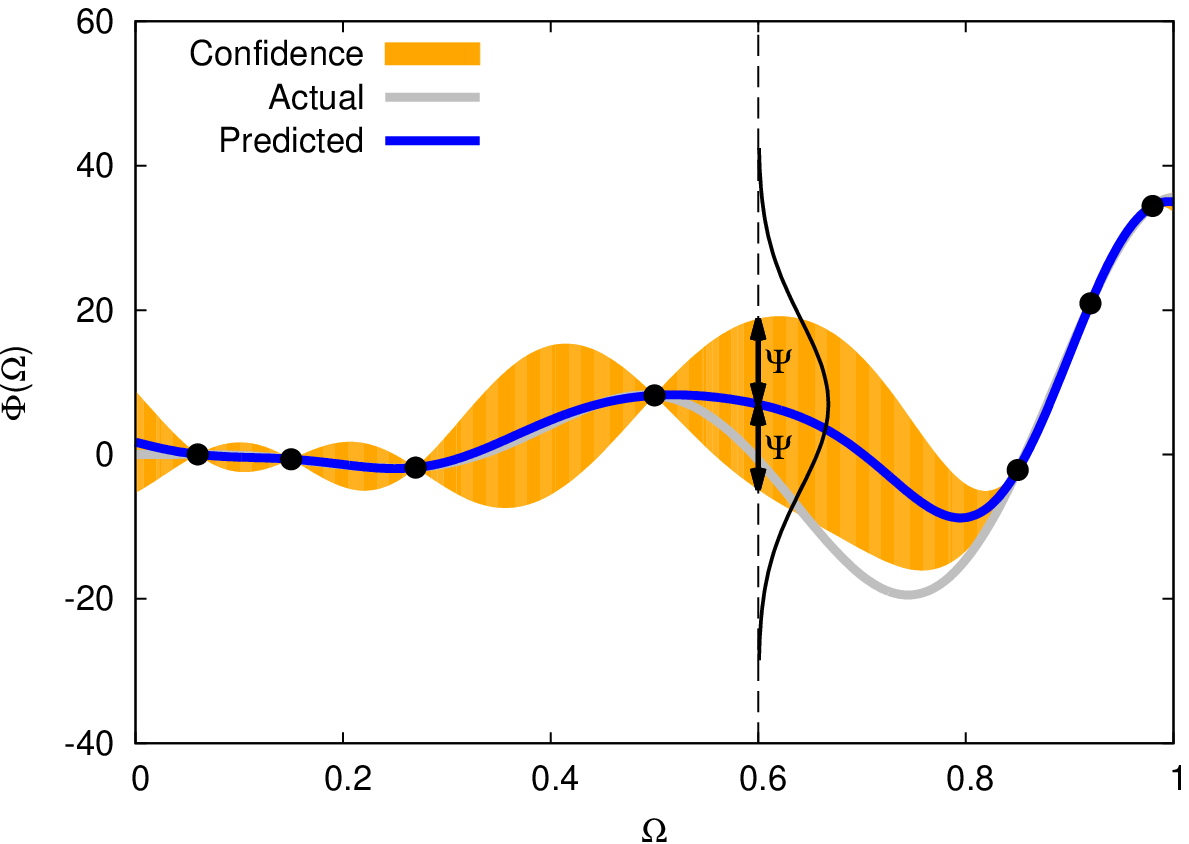}\tabularnewline
\end{tabular}\end{center}

\begin{center}~\vfill\end{center}

\begin{center}\textbf{Fig. 5 - A. Massa et} \textbf{\emph{al.}}\textbf{,}
\textbf{\emph{{}``}}On the Design of Complex ...''\end{center}
\newpage

\begin{center}\begin{tabular}{c}
\includegraphics[%
  width=0.63\columnwidth]{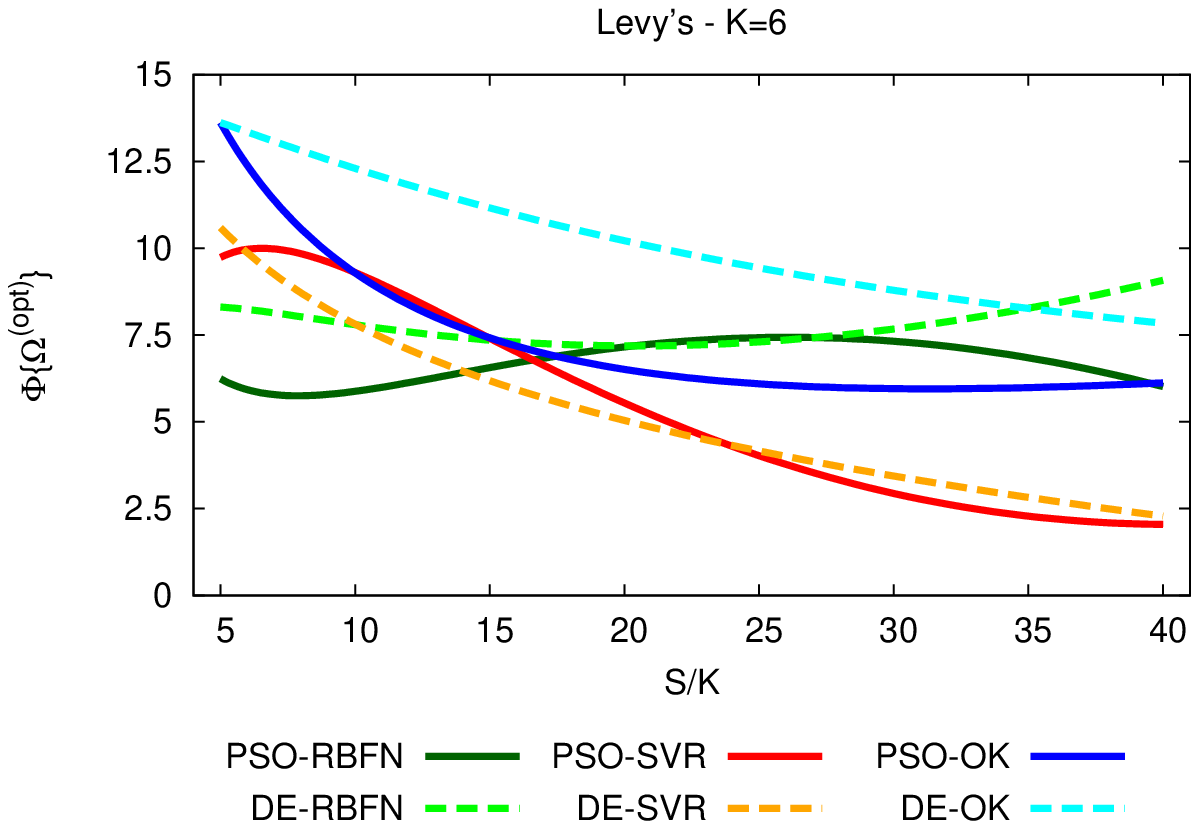}\tabularnewline
(\emph{a})\tabularnewline
\includegraphics[%
  width=0.63\columnwidth]{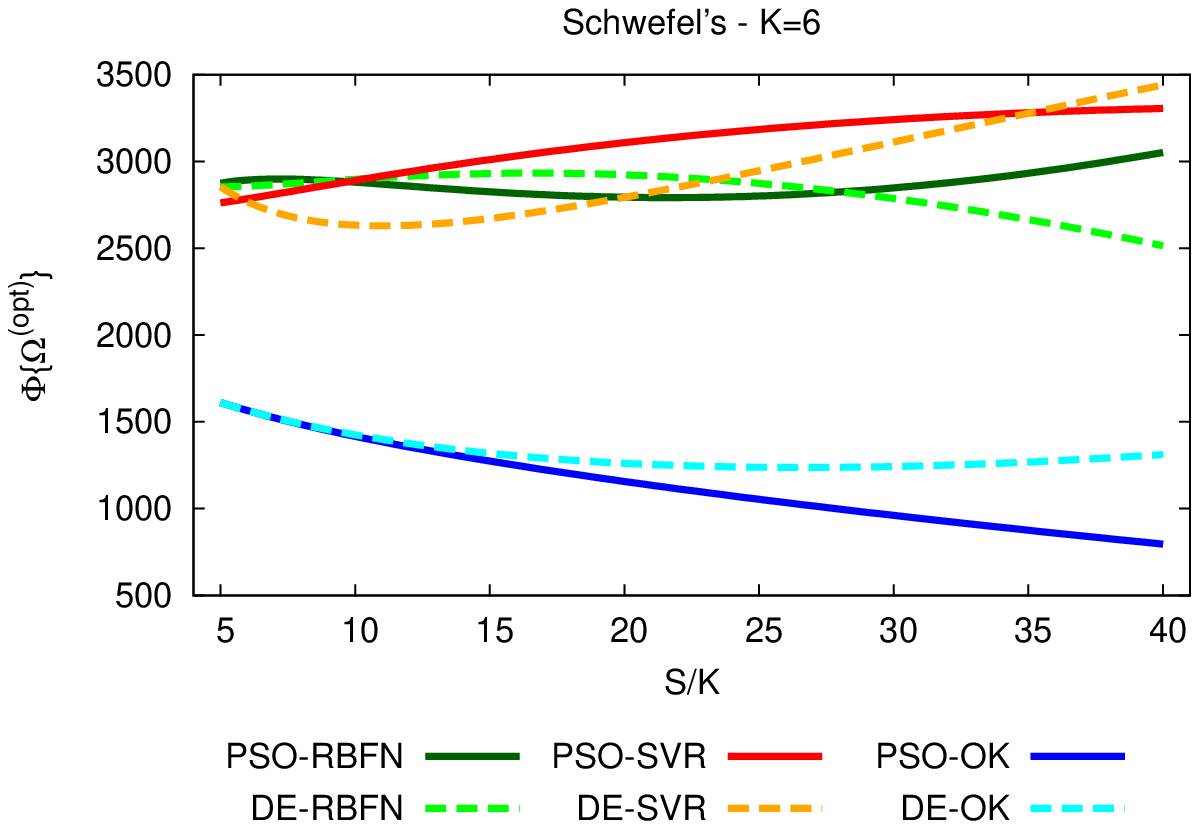}\tabularnewline
(\emph{b})\tabularnewline
\includegraphics[%
  width=0.63\columnwidth]{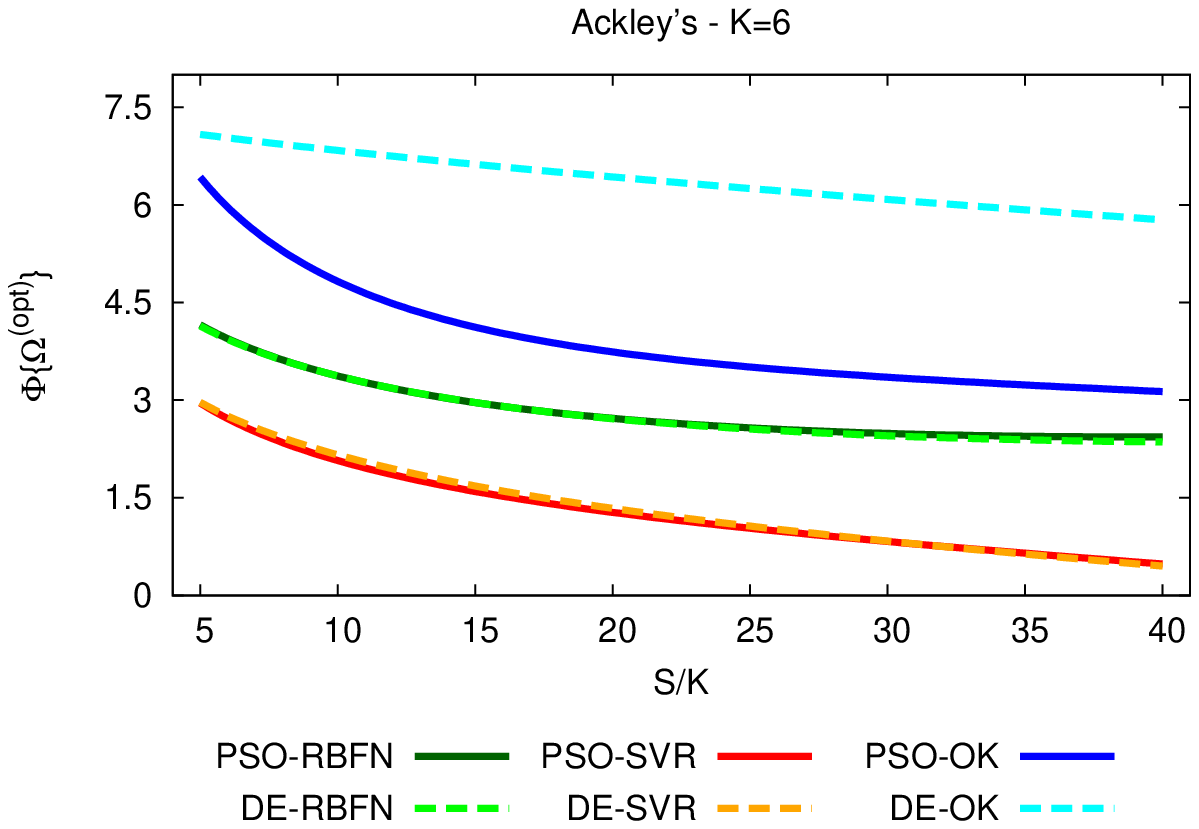}\tabularnewline
(\emph{c})\tabularnewline
\end{tabular}\end{center}

\begin{center}\textbf{Fig. 6 - A. Massa et} \textbf{\emph{al.}}\textbf{,}
\textbf{\emph{{}``}}On the Design of Complex ...''\end{center}
\newpage

\begin{center}~\vfill\end{center}

\begin{center}\begin{tabular}{c}
\includegraphics[%
  width=0.70\columnwidth]{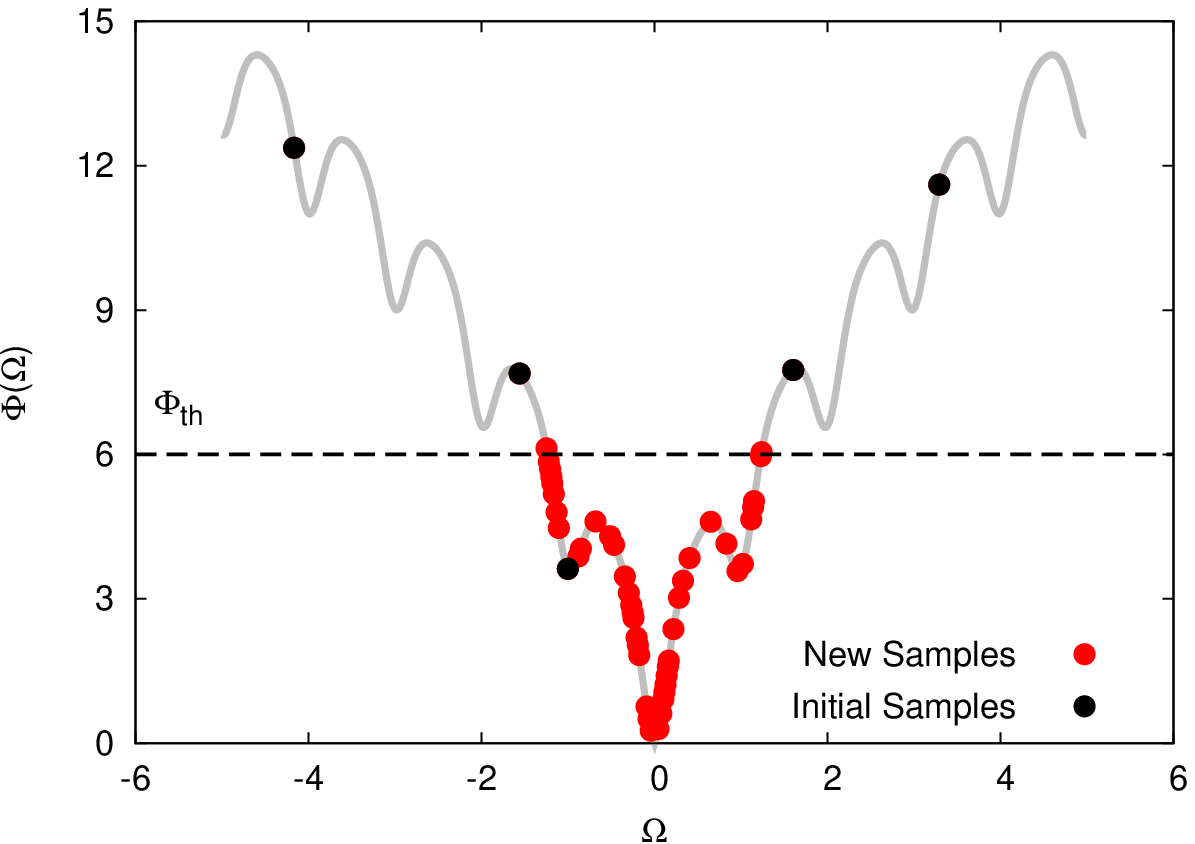}\tabularnewline
(\emph{a})\tabularnewline
\tabularnewline
\includegraphics[%
  width=0.70\columnwidth]{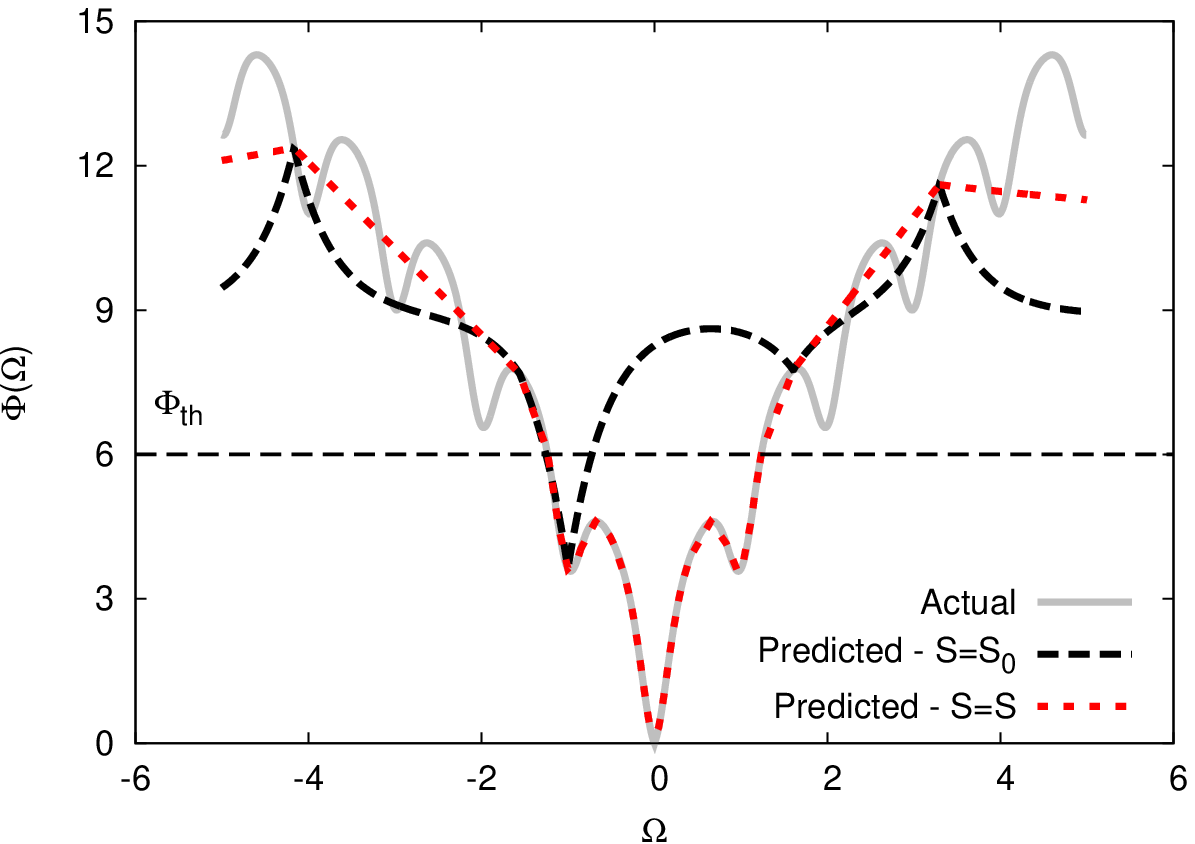}\tabularnewline
(\emph{b})\tabularnewline
\end{tabular}\end{center}

\begin{center}~\vfill\end{center}

\begin{center}\textbf{Fig. 7 - A. Massa et} \textbf{\emph{al.}}\textbf{,}
\textbf{\emph{{}``}}On the Design of Complex ...''\end{center}
\newpage

\begin{center}~\vfill\end{center}

\begin{center}\begin{tabular}{cc}
\includegraphics[%
  width=0.50\columnwidth]{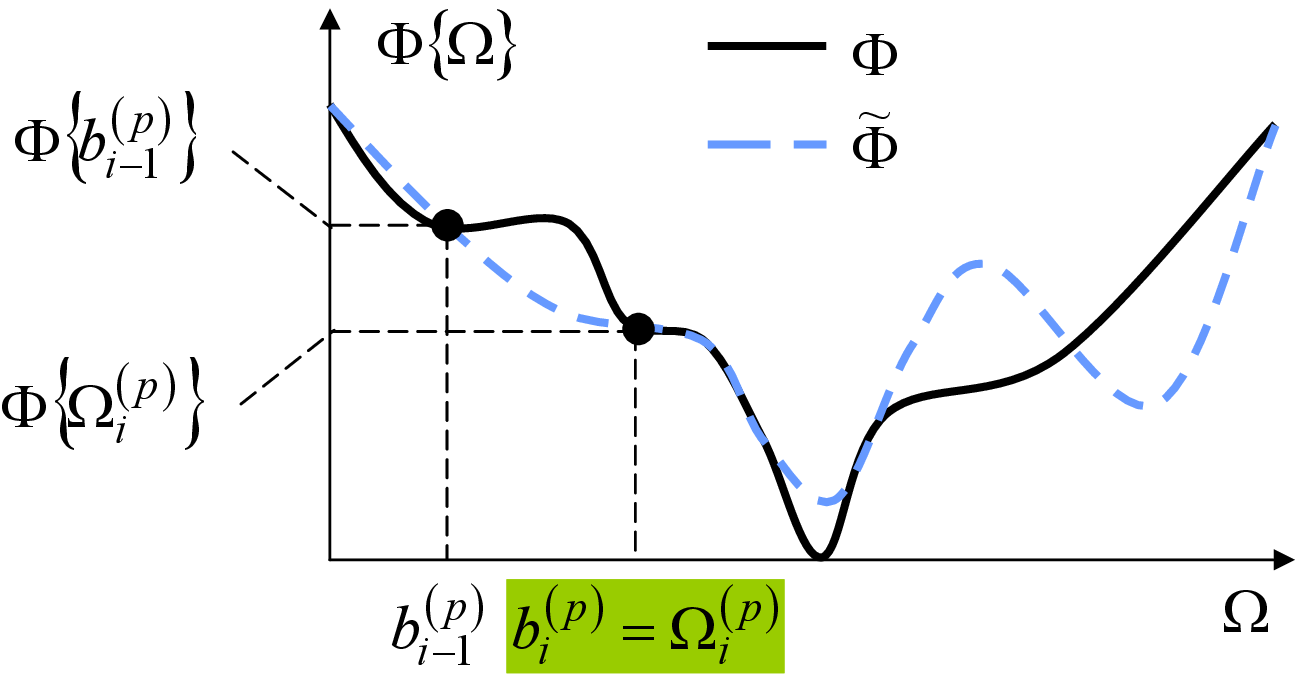}&
\includegraphics[%
  width=0.50\columnwidth]{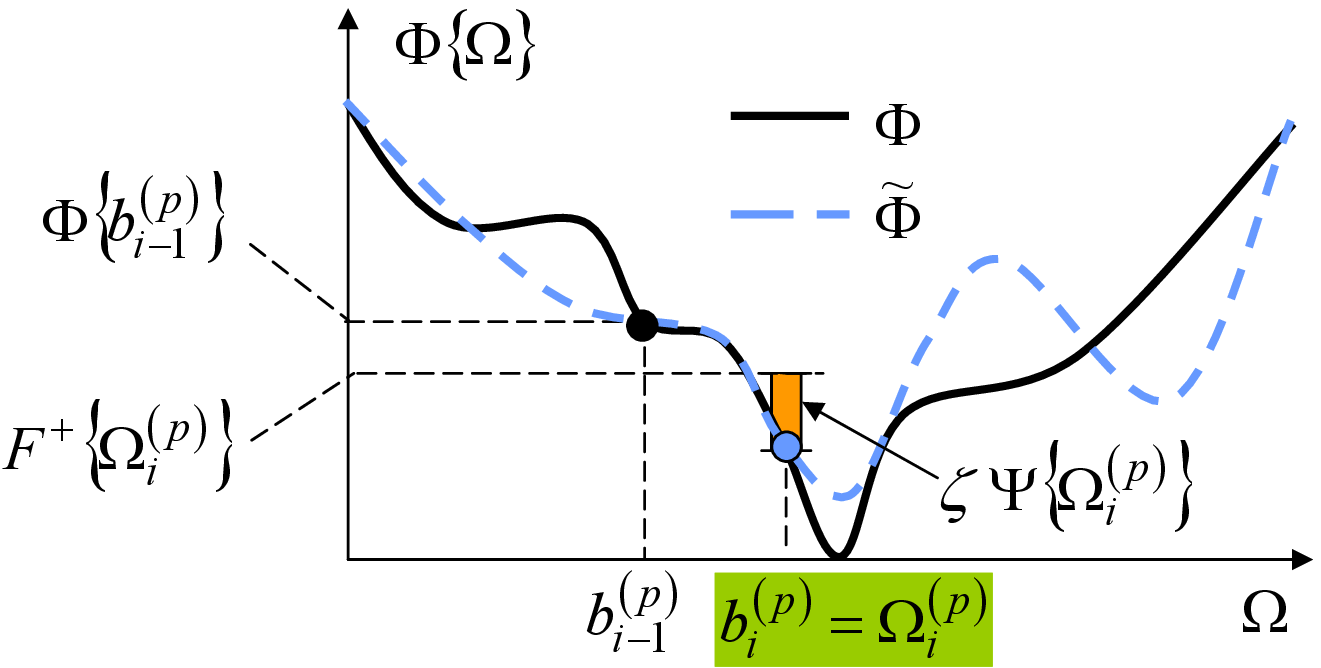}\tabularnewline
(\emph{a})&
(\emph{b})\tabularnewline
&
\tabularnewline
\includegraphics[%
  width=0.50\columnwidth]{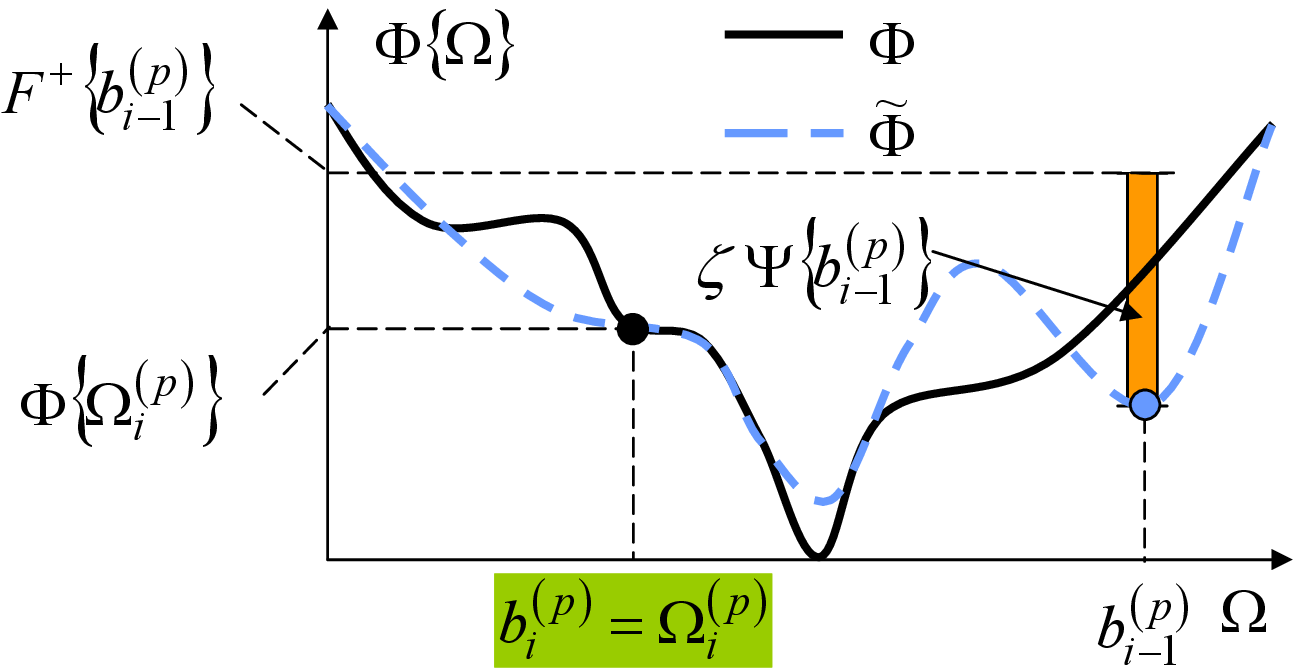}&
\includegraphics[%
  width=0.50\columnwidth]{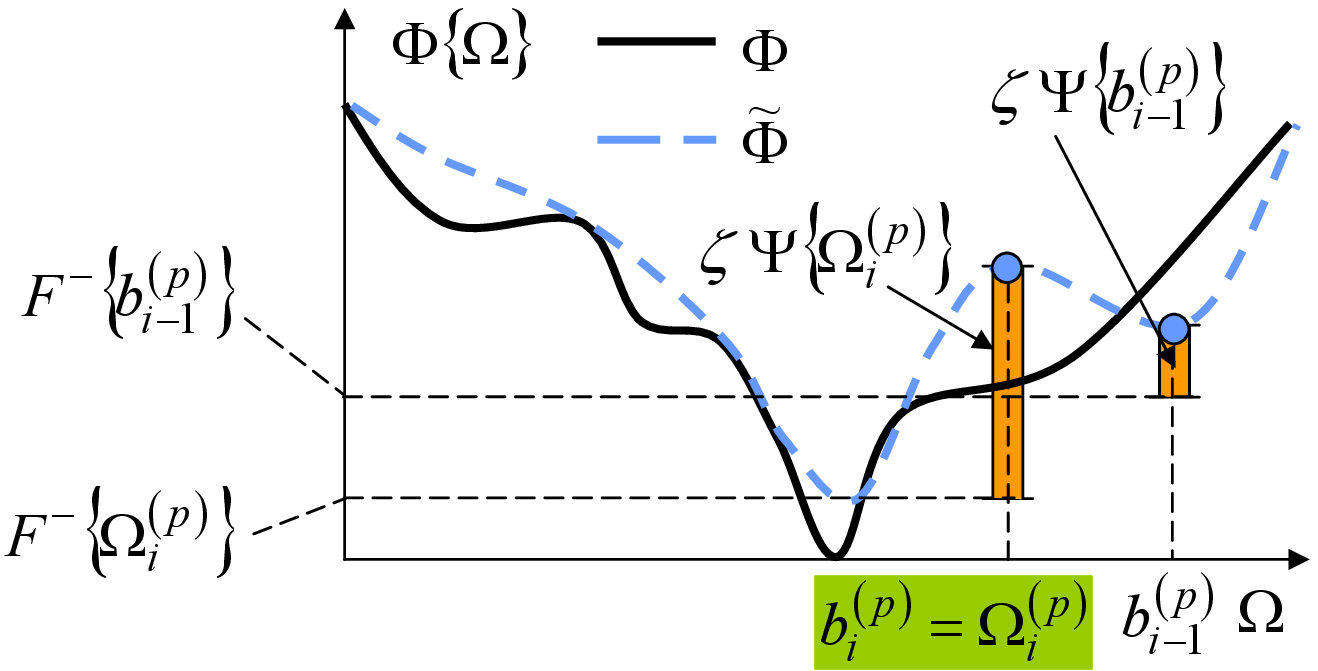}\tabularnewline
(\emph{c})&
(\emph{d})\tabularnewline
\end{tabular}\end{center}

\begin{center}~\vfill\end{center}

\begin{center}\textbf{Fig. 8 - A. Massa et} \textbf{\emph{al.}}\textbf{,}
\textbf{\emph{{}``}}On the Design of Complex ...''\end{center}
\newpage

\begin{center}~\vfill\end{center}

\begin{center}\begin{tabular}{c}
\includegraphics[%
  width=0.70\columnwidth]{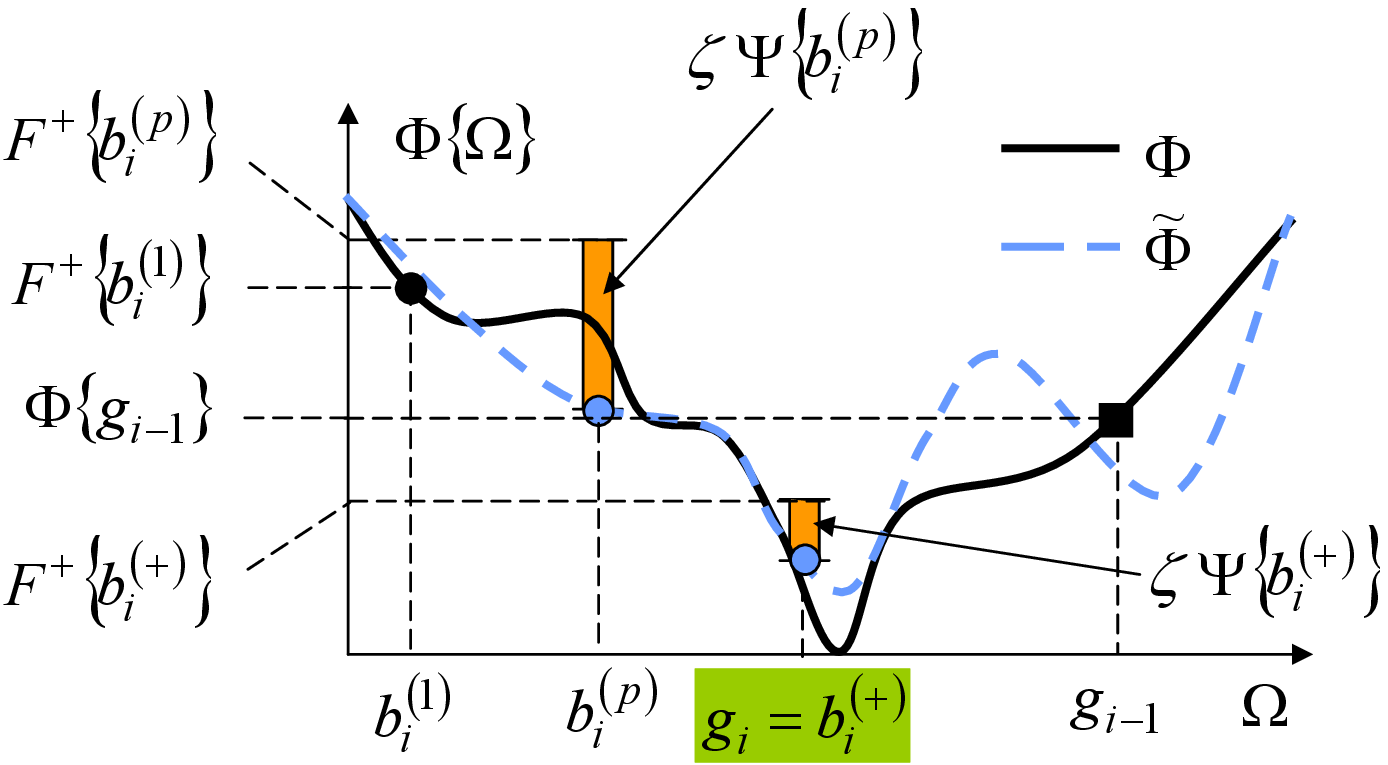}\tabularnewline
(\emph{a})\tabularnewline
\tabularnewline
\includegraphics[%
  width=0.70\columnwidth]{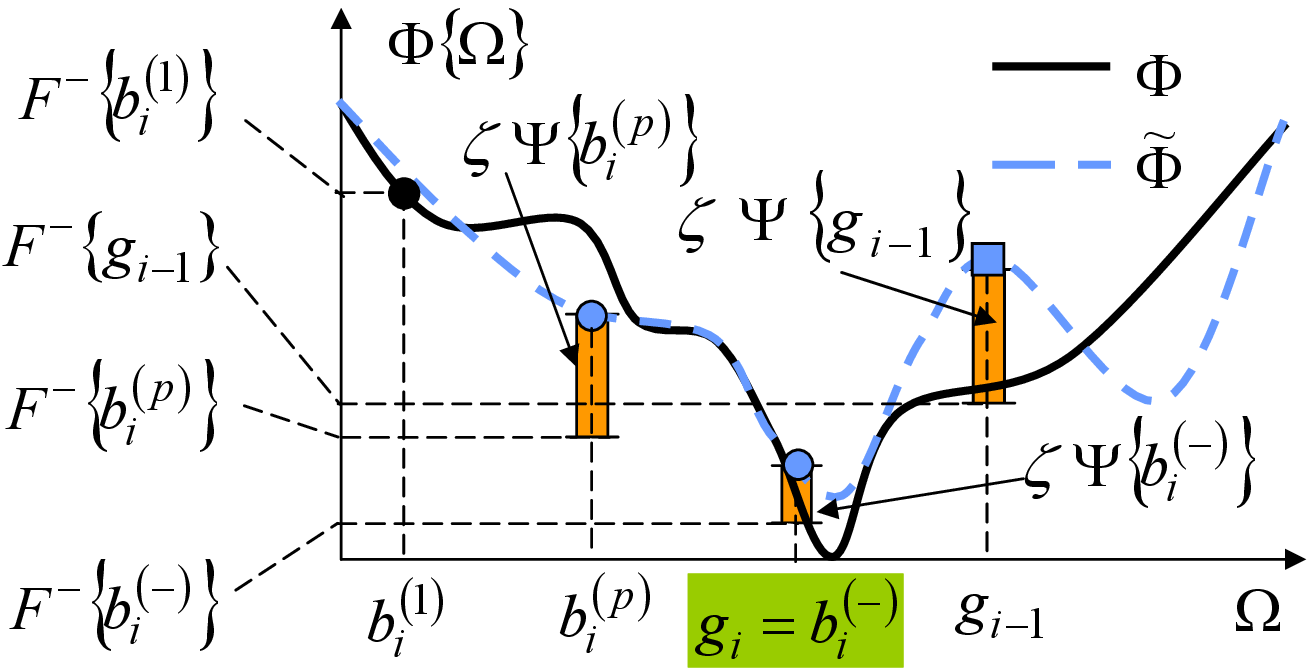}\tabularnewline
(\emph{b})\tabularnewline
\end{tabular}\end{center}

\begin{center}~\vfill\end{center}

\begin{center}\textbf{Fig. 9 - A. Massa et} \textbf{\emph{al.}}\textbf{,}
\textbf{\emph{{}``}}On the Design of Complex ...''\end{center}
\newpage

\begin{center}\begin{tabular}{c}
\includegraphics[%
  width=0.63\columnwidth]{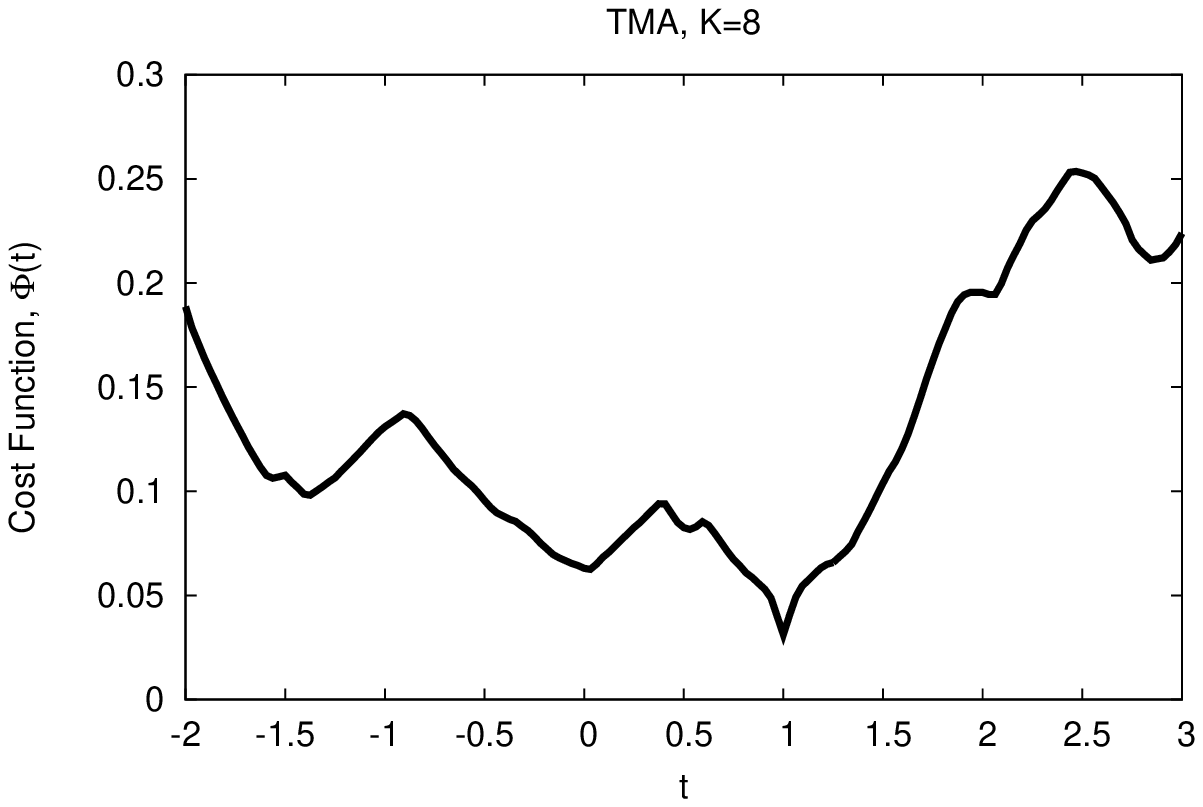}\tabularnewline
(\emph{a})\tabularnewline
\includegraphics[%
  width=0.63\columnwidth]{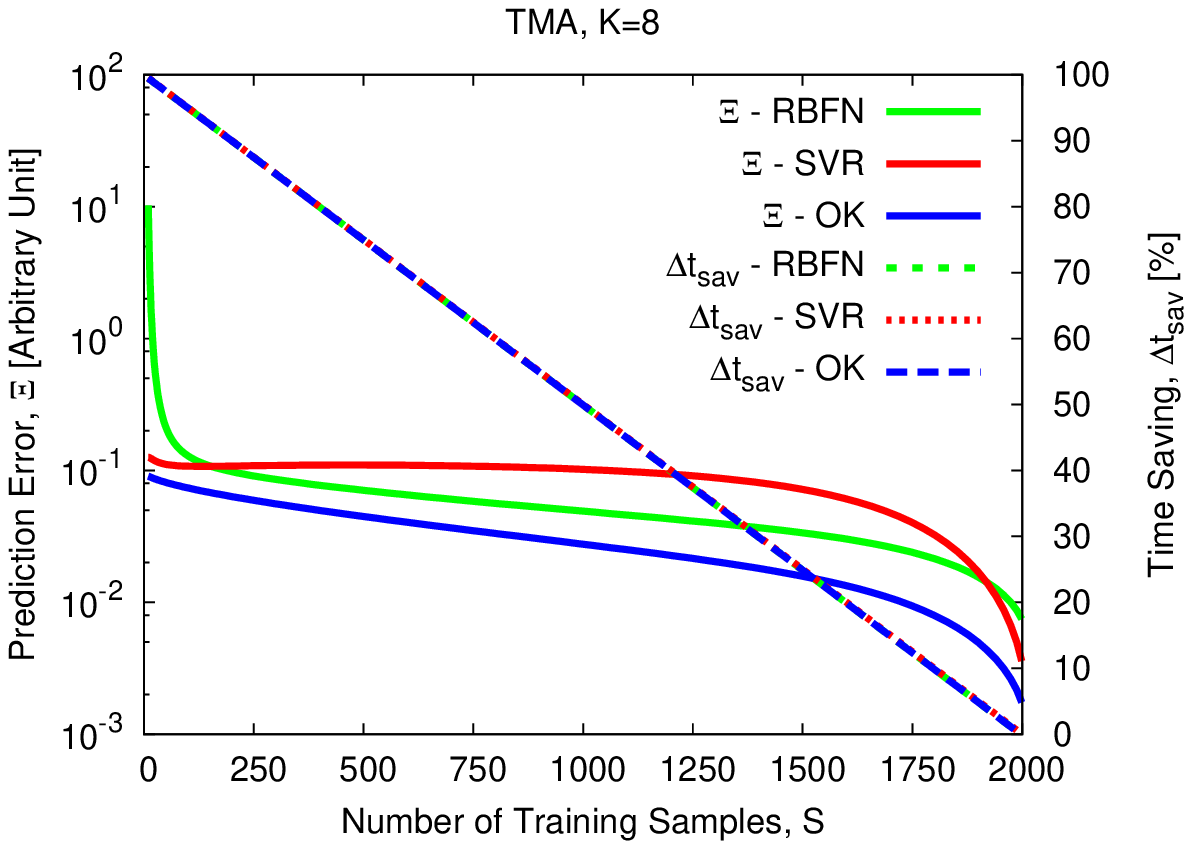}\tabularnewline
(\emph{b})\tabularnewline
\includegraphics[%
  width=0.63\columnwidth]{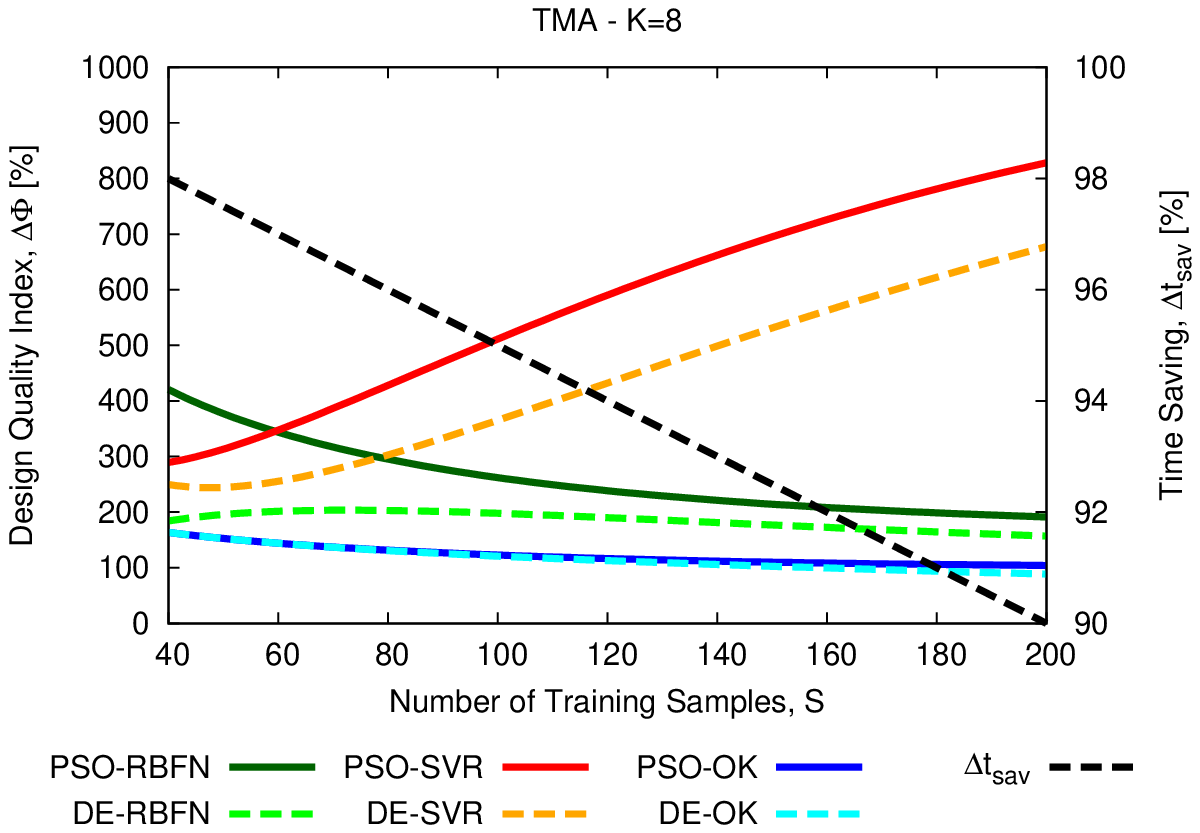}\tabularnewline
(\emph{c})\tabularnewline
\end{tabular}\end{center}

\begin{center}\textbf{Fig. 10 - A. Massa et} \textbf{\emph{al.}}\textbf{,}
\textbf{\emph{{}``}}On the Design of Complex ...''\end{center}
\newpage

\begin{center}~\vfill\end{center}

\begin{center}\begin{tabular}{c}
\includegraphics[%
  width=1.0\columnwidth]{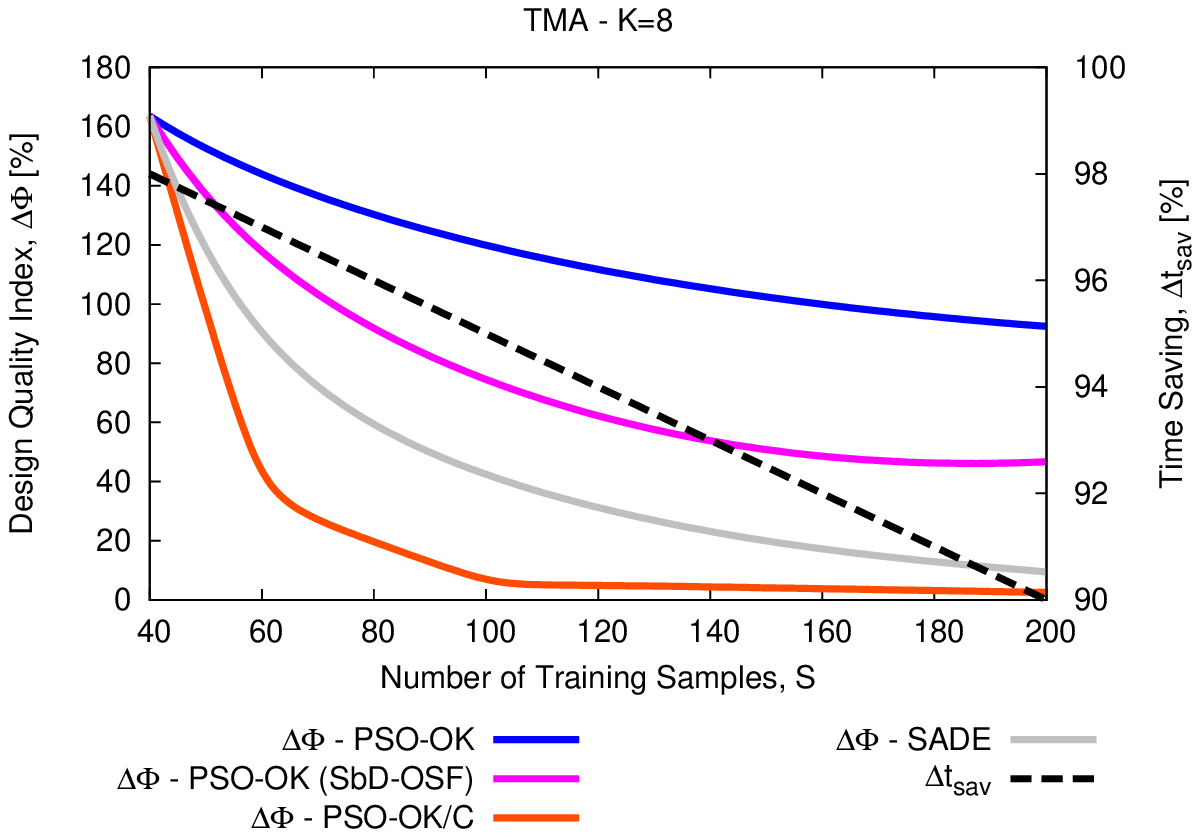}\tabularnewline
\end{tabular}\end{center}

\begin{center}~\vfill\end{center}

\begin{center}\textbf{Fig. 11 - A. Massa et} \textbf{\emph{al.}}\textbf{,}
\textbf{\emph{{}``}}On the Design of Complex ...''\end{center}
\newpage

\begin{center}~\vfill\end{center}

\begin{center}\begin{tabular}{c}
\includegraphics[%
  width=0.70\columnwidth]{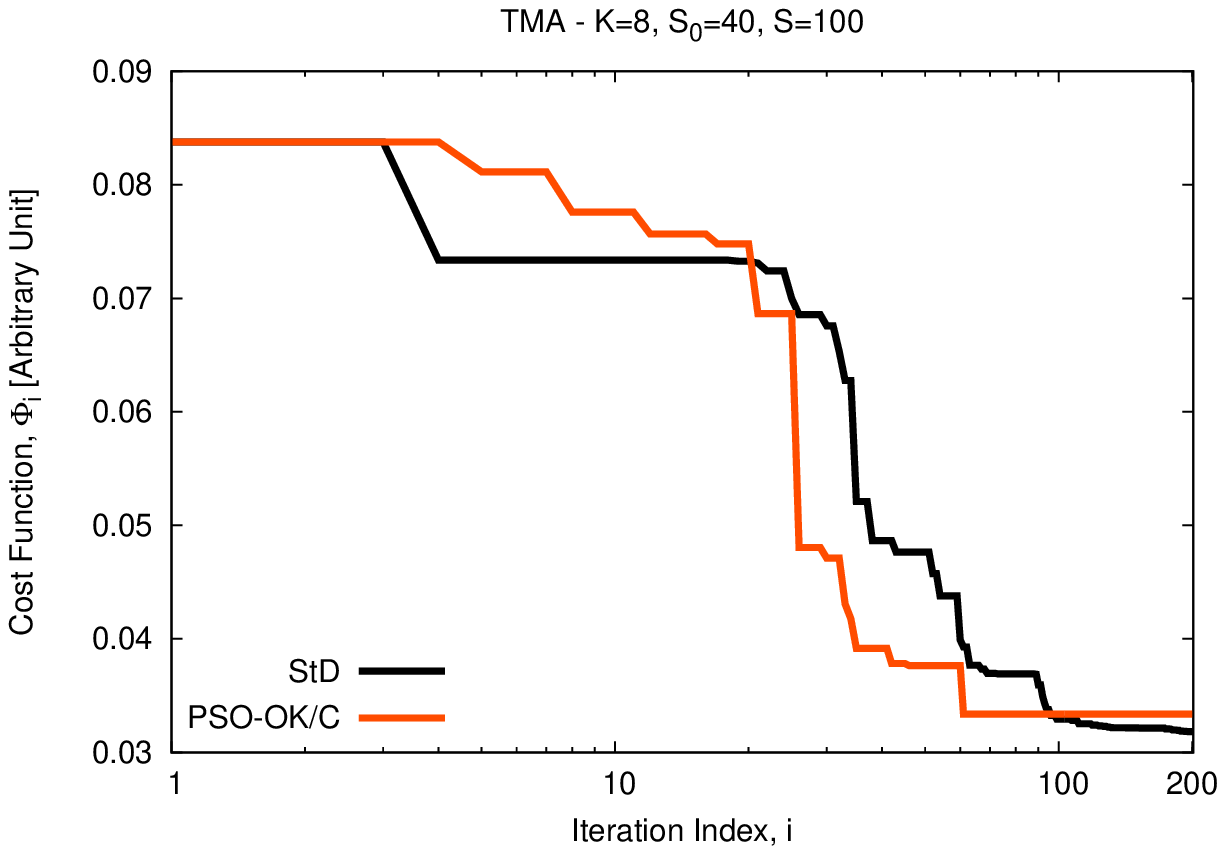}\tabularnewline
(\emph{a})\tabularnewline
\tabularnewline
\includegraphics[%
  width=0.70\columnwidth]{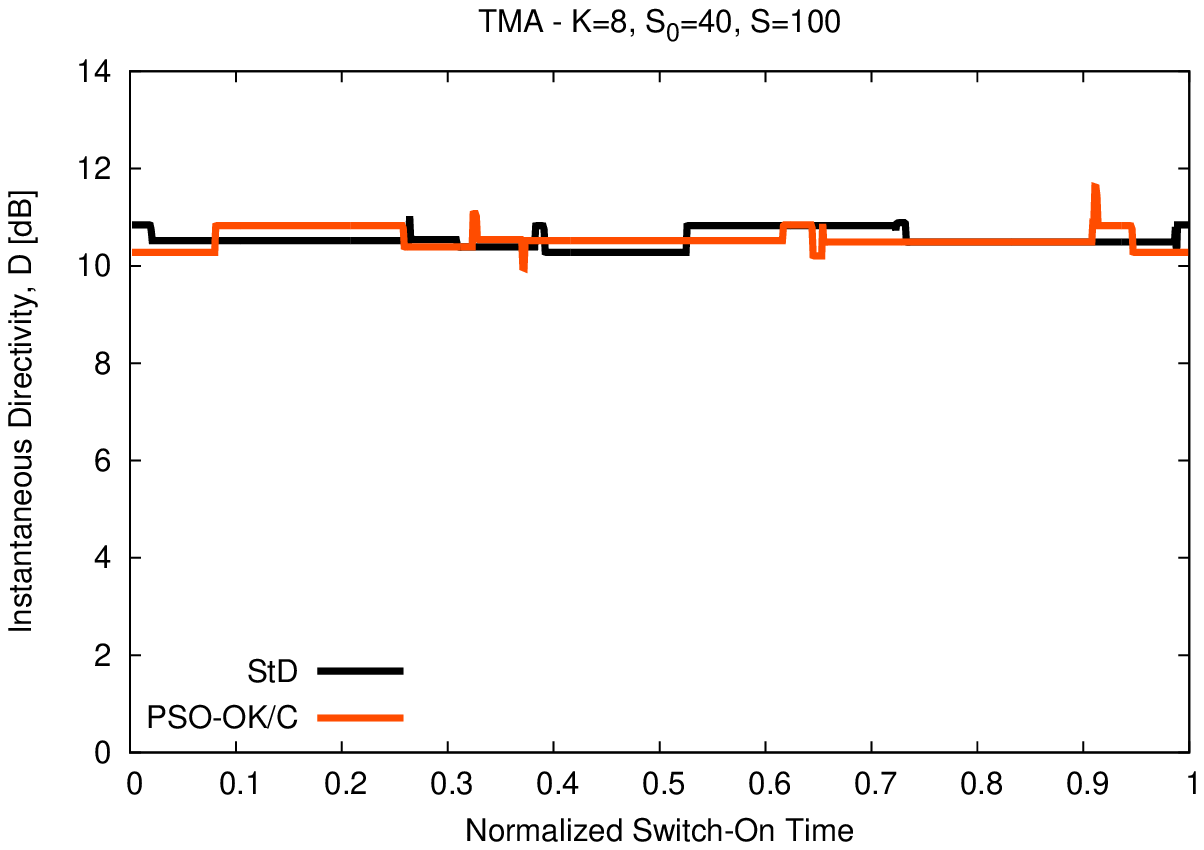}\tabularnewline
(\emph{b})\tabularnewline
\end{tabular}\end{center}

\begin{center}~\vfill\end{center}

\begin{center}\textbf{Fig. 12 - A. Massa et} \textbf{\emph{al.}}\textbf{,}
\textbf{\emph{{}``}}On the Design of Complex ...''\end{center}
\newpage

\begin{center}~\vfill\end{center}

\begin{center}\begin{tabular}{c}
\includegraphics[%
  width=1.0\columnwidth]{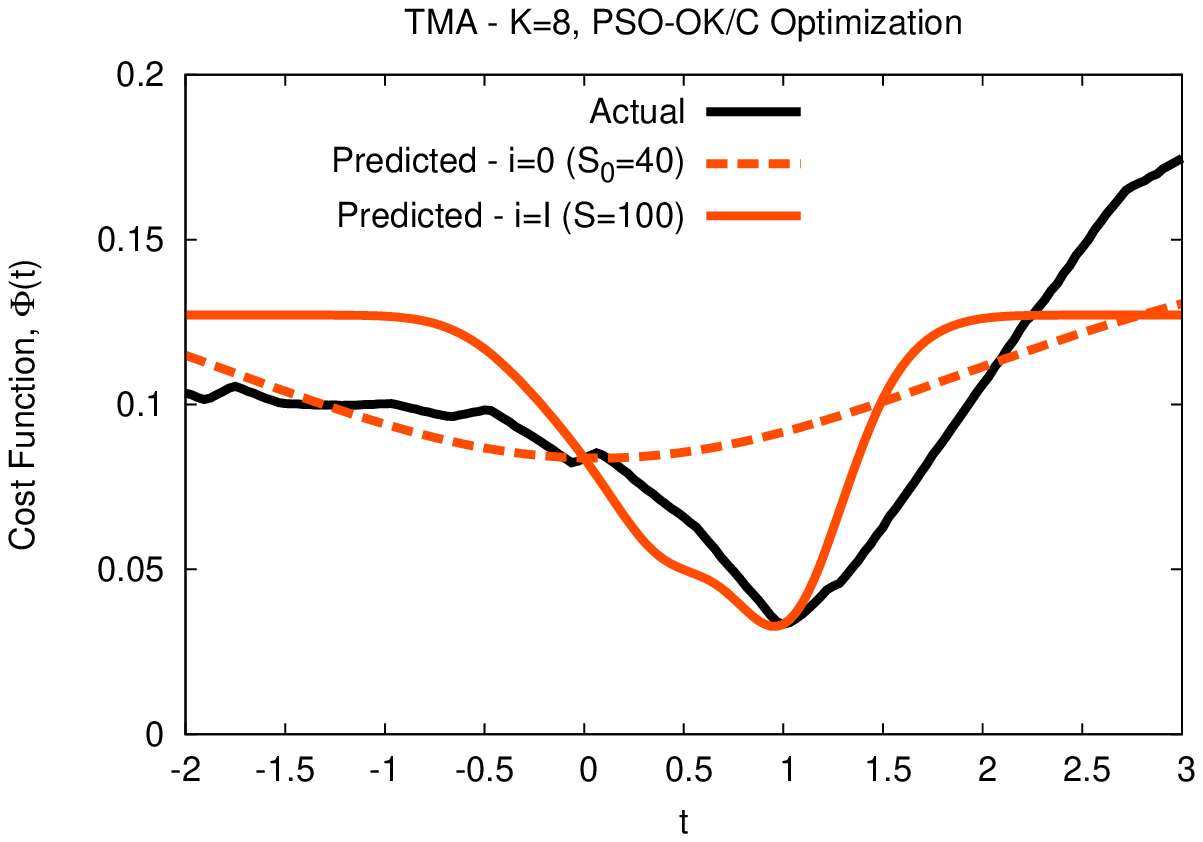}\tabularnewline
\end{tabular}\end{center}

\begin{center}~\vfill\end{center}

\begin{center}\textbf{Fig. 13 - A. Massa et} \textbf{\emph{al.}}\textbf{,}
\textbf{\emph{{}``}}On the Design of Complex ...''\end{center}
\newpage

\begin{center}~\vfill\end{center}

\begin{center}\begin{tabular}{c}
\includegraphics[%
  width=1.0\columnwidth]{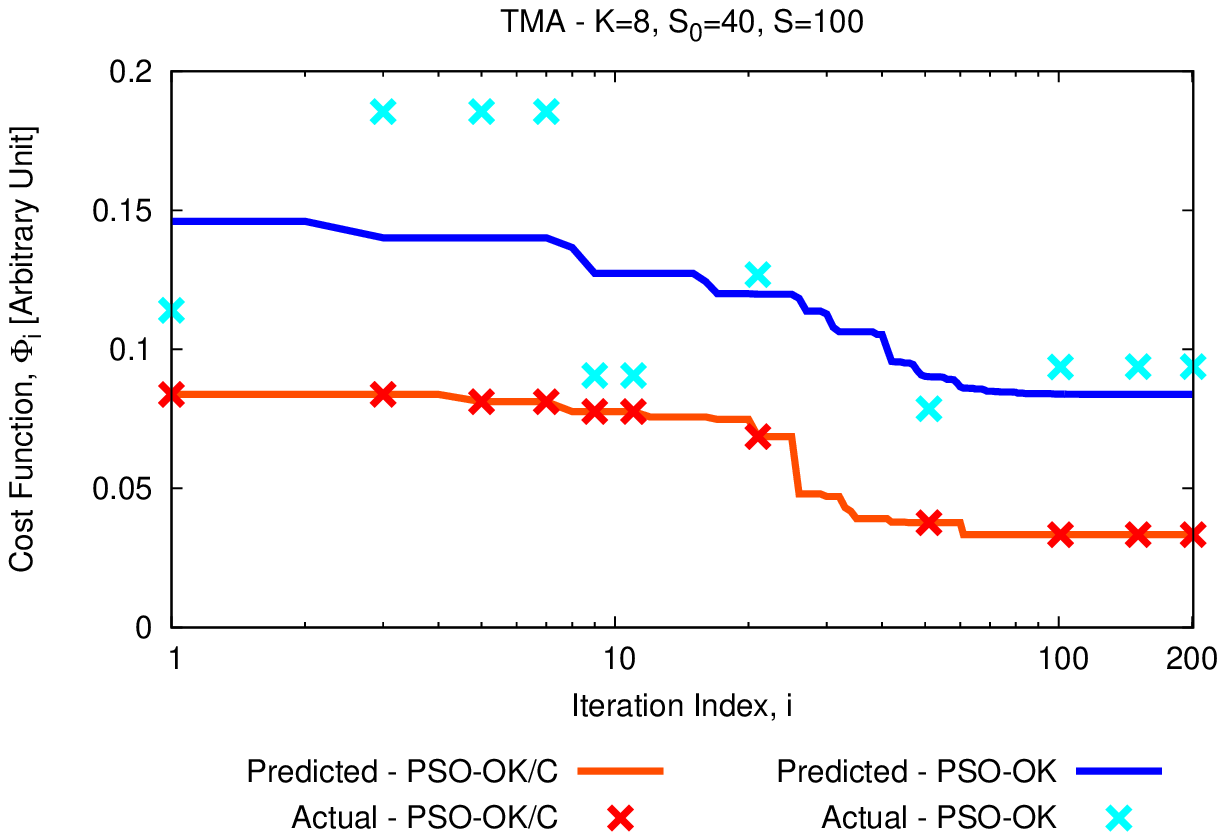}\tabularnewline
\end{tabular}\end{center}

\begin{center}~\vfill\end{center}

\begin{center}\textbf{Fig. 14 - A. Massa et} \textbf{\emph{al.}}\textbf{,}
\textbf{\emph{{}``}}On the Design of Complex ...''\end{center}
\newpage

\begin{center}~\vfill\end{center}

\begin{center}\begin{tabular}{c}
\includegraphics[%
  width=0.65\columnwidth]{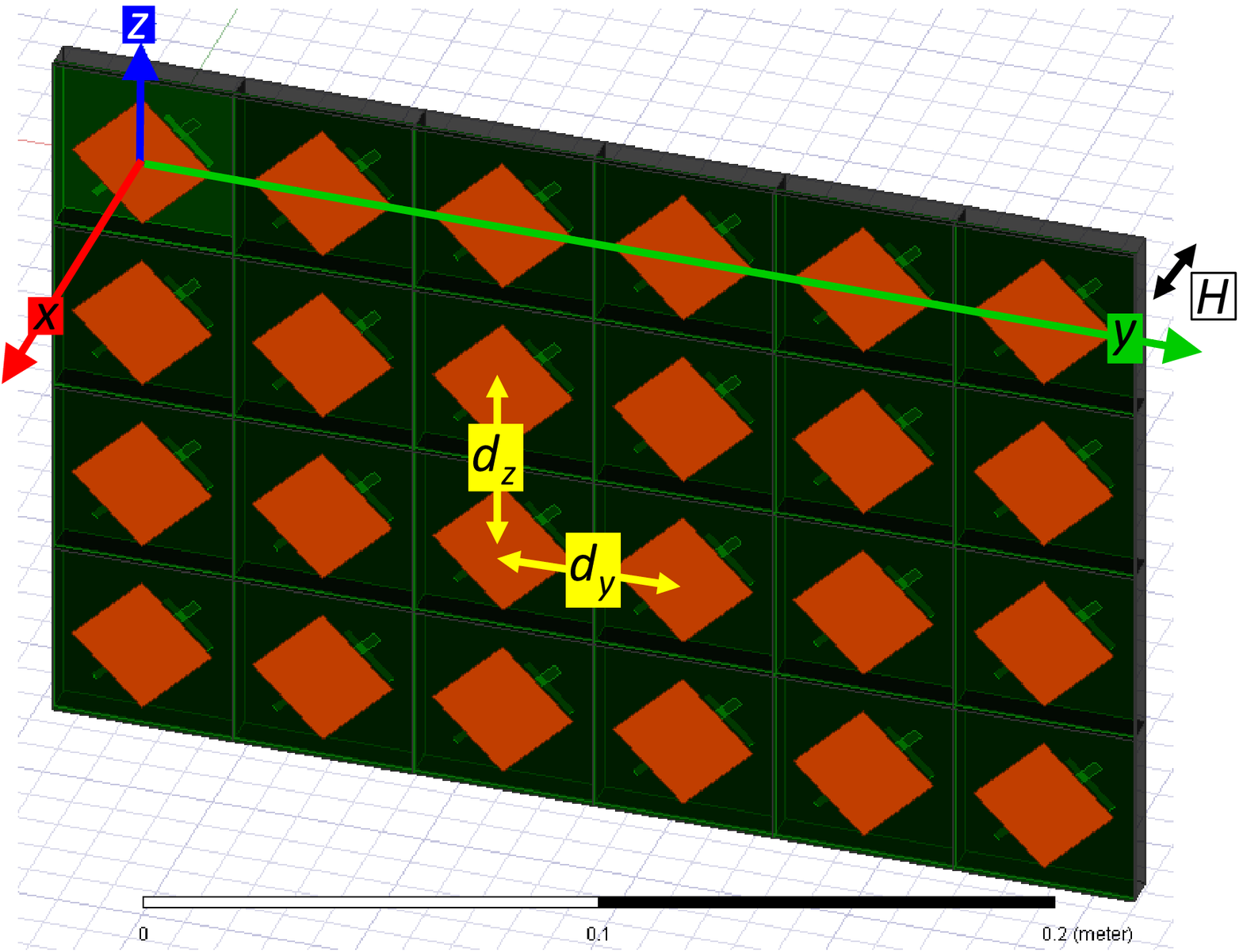}\tabularnewline
(\emph{a})\tabularnewline
\tabularnewline
\includegraphics[%
  width=0.65\columnwidth]{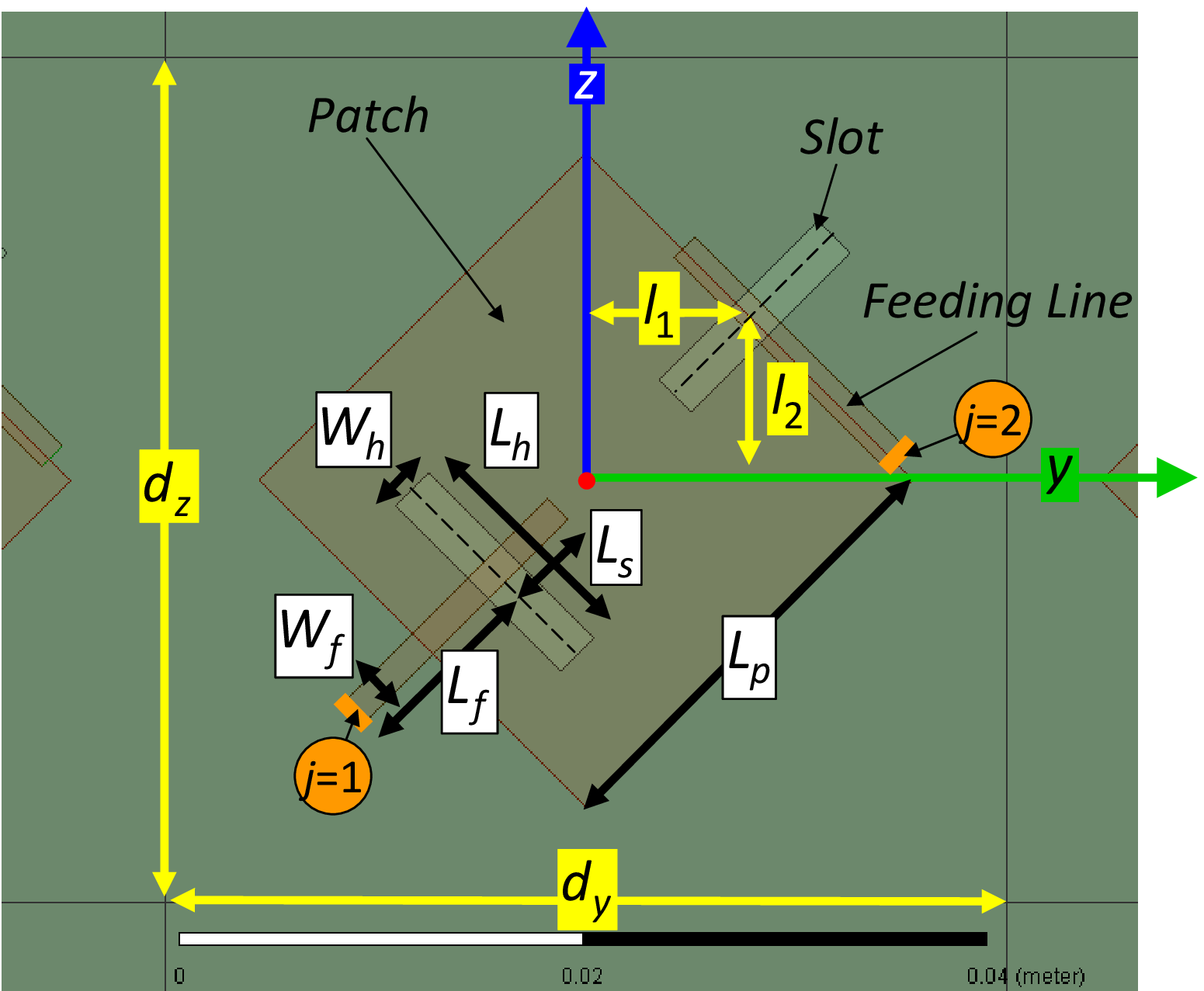}\tabularnewline
(\emph{b})\tabularnewline
\end{tabular}\end{center}

\begin{center}~\vfill\end{center}

\begin{center}\textbf{Fig. 15 - A. Massa et} \textbf{\emph{al.}}\textbf{,}
\textbf{\emph{{}``}}On the Design of Complex ...''\end{center}
\newpage

\begin{center}~\vfill\end{center}

\begin{center}\begin{tabular}{c}
\includegraphics[%
  width=1.0\columnwidth]{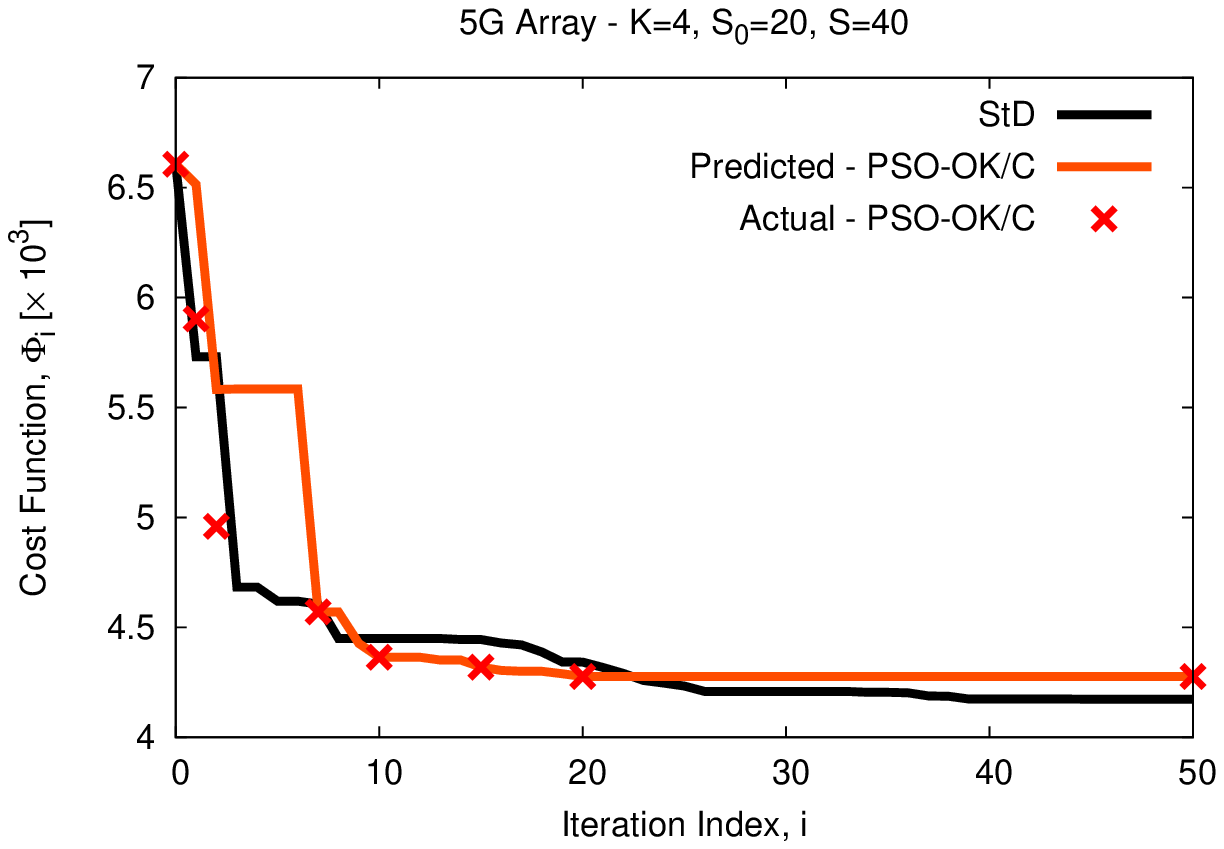}\tabularnewline
\end{tabular}\end{center}

\begin{center}~\vfill\end{center}

\begin{center}\textbf{Fig. 16 - A. Massa et} \textbf{\emph{al.}}\textbf{,}
\textbf{\emph{{}``}}On the Design of Complex ...''\end{center}
\newpage

\begin{center}~\vfill\end{center}

\begin{center}\begin{tabular}{c}
\includegraphics[%
  width=0.70\columnwidth]{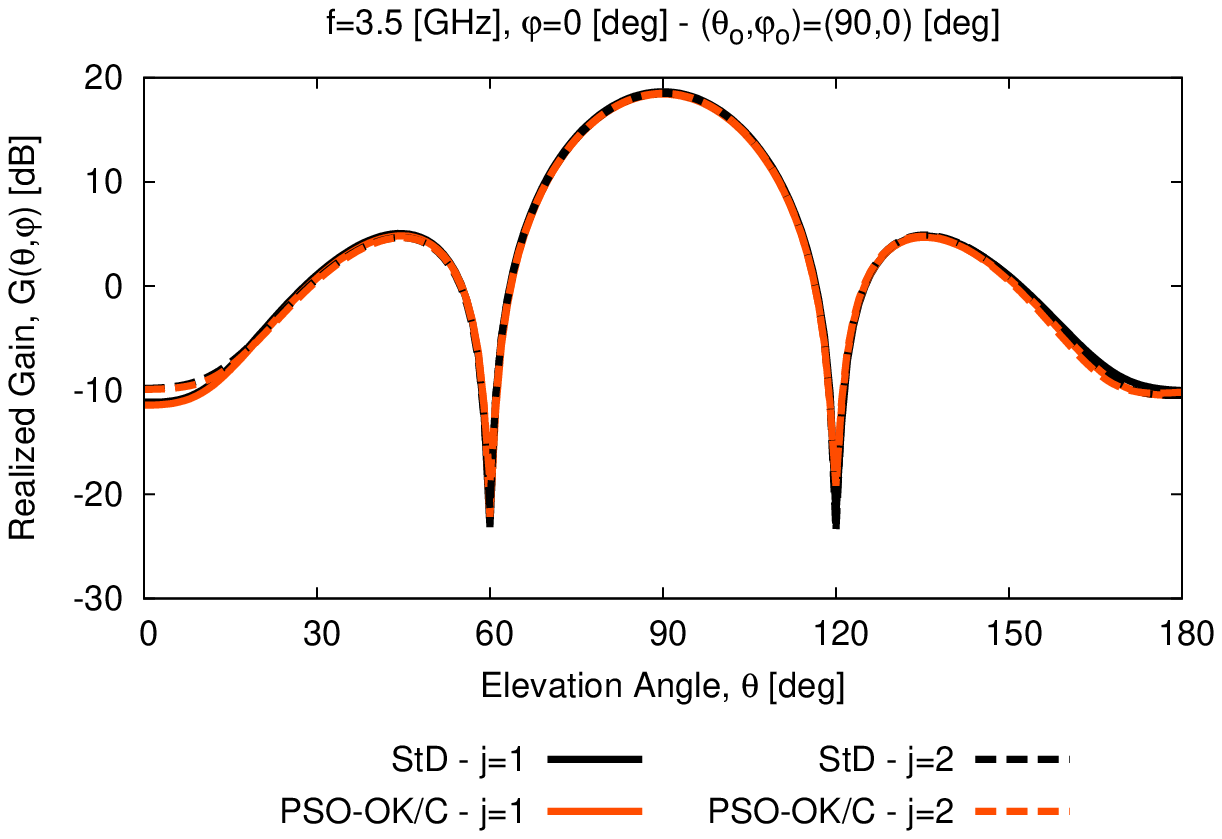}\tabularnewline
(\emph{a})\tabularnewline
\tabularnewline
\includegraphics[%
  width=0.70\columnwidth]{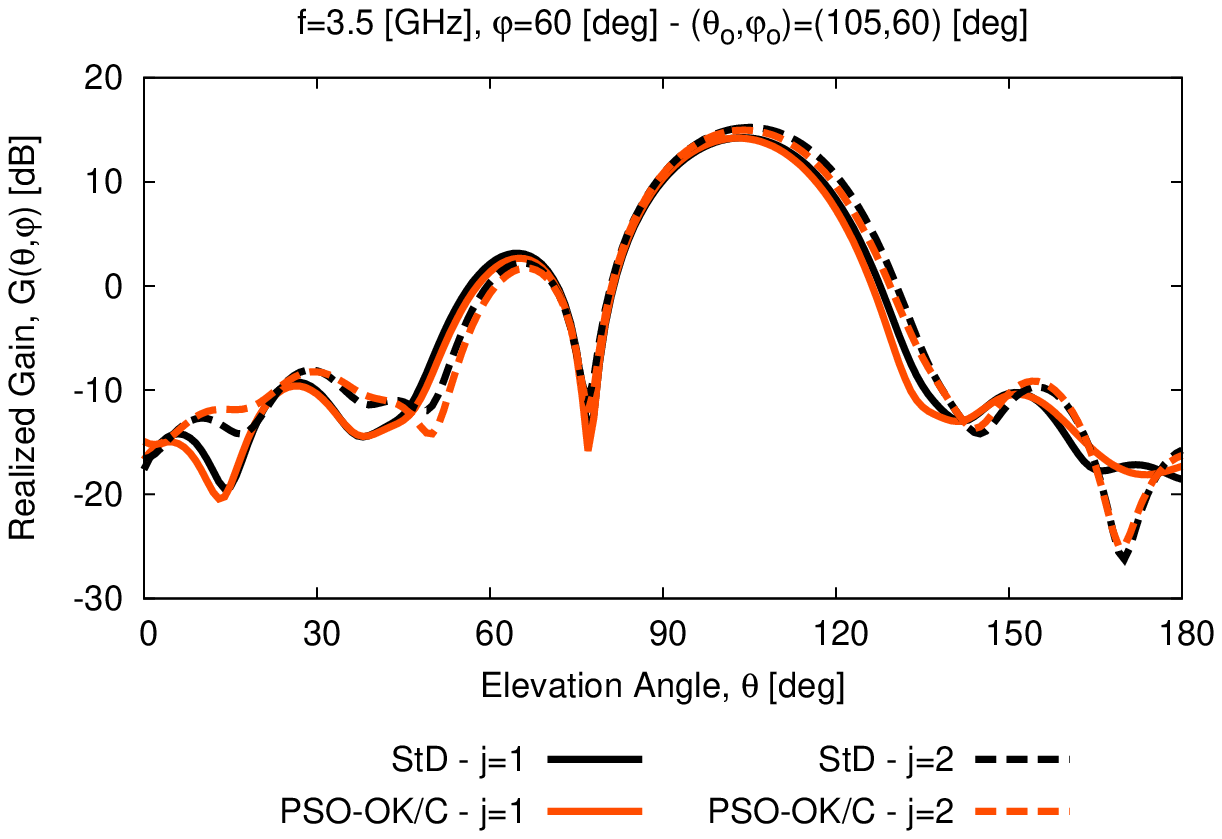}\tabularnewline
(\emph{b})\tabularnewline
\end{tabular}\end{center}

\begin{center}~\vfill\end{center}

\begin{center}\textbf{Fig. 17 - A. Massa et} \textbf{\emph{al.}}\textbf{,}
\textbf{\emph{{}``}}On the Design of Complex ...''\end{center}
\newpage

\begin{center}~\vfill\end{center}

\begin{center}\begin{tabular}{|c||c|c|}
\hline 
&
{\scriptsize $\underline{\Omega}_{i}^{\left(p\right)}$ Simulated}&
{\scriptsize $\underline{\Omega}_{i}^{\left(p\right)}$ Predicted}\tabularnewline
\hline
\hline 
{\scriptsize $\underline{b}_{i-1}^{\left(p\right)}$ Simulated}&
{\scriptsize $\underline{b}_{i}^{\left(p\right)}=\left\{ \begin{array}{ll}
\underline{\Omega}_{i}^{\left(p\right)} & \mathrm{{if}}\,\Phi\left\{ \underline{\Omega}_{i}^{\left(p\right)}\right\} <\Phi\left\{ \underline{b}_{i-1}^{\left(p\right)}\right\} \\
\underline{b}_{i-1}^{\left(p\right)} & \mathrm{{otherwise}}\end{array}\right.$}&
{\scriptsize $\underline{b}_{i}^{\left(p\right)}=\left\{ \begin{array}{ll}
\underline{\Omega}_{i}^{\left(p\right)} & \mathrm{{if}}\,\mathcal{F}^{+}\left\{ \underline{\Omega}_{i}^{\left(p\right)}\right\} <\Phi\left\{ \underline{b}_{i-1}^{\left(p\right)}\right\} \\
\underline{b}_{i-1}^{\left(p\right)} & \mathrm{{otherwise}}\end{array}\right.$}\tabularnewline
\hline 
{\scriptsize $\underline{b}_{i-1}^{\left(p\right)}$ Predicted}&
{\scriptsize $\underline{b}_{i}^{\left(p\right)}=\left\{ \begin{array}{ll}
\underline{\Omega}_{i}^{\left(p\right)} & \mathrm{{if}}\,\Phi\left\{ \underline{\Omega}_{i}^{\left(p\right)}\right\} <\mathcal{F}^{+}\left\{ \underline{b}_{i-1}^{\left(p\right)}\right\} \\
\underline{b}_{i-1}^{\left(p\right)} & \mathrm{{otherwise}}\end{array}\right.$}&
{\scriptsize $\underline{b}_{i}^{\left(p\right)}=\left\{ \begin{array}{ll}
\underline{\Omega}_{i}^{\left(p\right)} & \mathrm{{if}}\,\mathcal{F}^{-}\left\{ \underline{\Omega}_{i}^{\left(p\right)}\right\} <\mathcal{F}^{-}\left\{ \underline{b}_{i-1}^{\left(p\right)}\right\} \\
\underline{b}_{i-1}^{\left(p\right)} & \mathrm{{otherwise}}\end{array}\right.$}\tabularnewline
\hline
\end{tabular}\end{center}

\begin{center}~\vfill\end{center}

\begin{center}\textbf{Tab. I - A. Massa et} \textbf{\emph{al.}}\textbf{,}
\textbf{\emph{{}``}}On the Design of Complex ...''\end{center}
\newpage

\begin{center}~\vfill\end{center}

\begin{center}\begin{tabular}{|c||c|}
\hline 
$\underline{g}_{i-1}$ Simulated&
$\begin{array}{l}
\underline{g}_{i}=\left\{ \begin{array}{ll}
\underline{b}_{i}^{\left(+\right)} & \mathrm{{if}}\,\mathcal{F}^{+}\left\{ \underline{b}_{i}^{\left(+\right)}\right\} <\Phi\left\{ \underline{g}_{i-1}\right\} \\
\underline{g}_{i-1} & \mathrm{{otherwise}}\end{array}\right.\\
\mathrm{{s.t.}}\,\,\underline{b}_{i}^{\left(+\right)}=\arg\left[\min_{p=1,\,...,\, P}\mathcal{F}^{+}\left\{ \underline{b}_{i}^{\left(p\right)}\right\} \right]\end{array}$\tabularnewline
\hline 
$\underline{g}_{i-1}$ Predicted&
$\begin{array}{l}
\underline{g}_{i}=\left\{ \begin{array}{ll}
\underline{b}_{i}^{\left(-\right)} & \mathrm{{if}}\,\mathcal{F}^{-}\left\{ \underline{b}_{i}^{\left(-\right)}\right\} <\mathcal{F}^{-}\left\{ \underline{g}_{i-1}\right\} \\
\underline{g}_{i-1} & \mathrm{{otherwise}}\end{array}\right.\\
\mathrm{{s.t.}}\,\,\underline{b}_{i}^{\left(-\right)}=\arg\left[\min_{p=1,\,...,\, P}\mathcal{F}^{-}\left\{ \underline{b}_{i}^{\left(p\right)}\right\} \right]\end{array}$\tabularnewline
\hline
\end{tabular} \end{center}

\begin{center}~\vfill\end{center}

\begin{center}\textbf{Tab. II - A. Massa et} \textbf{\emph{al.}}\textbf{,}
\textbf{\emph{{}``}}On the Design of Complex ...''\end{center}
\end{document}